%% file: 00main_jwst_sgrb2.tex
\documentclass[twocolumn]{aastex701}

\usepackage{graphicx}
\input LaTeX-macros.tex
\usepackage[utf8]{inputenc}
\usepackage{xspace}
\usepackage{natbib}
\usepackage{array}
\usepackage{subcaption}

\newcommand{\rr}[1]{#1}
\newcommand{\rrr}[1]{#1}
\newcommand{\rf}[1]{#1}

\begin{document} 

\title{JWST's first view of the most vigorously star-forming cloud in the Galactic center -- Sagittarius B2}

\author[0000-0002-0533-8575]{Nazar Budaiev}
\affiliation{Department of Astronomy, University of Florida, P.O. Box 112055, Gainesville, FL 32611, USA}
\email[show]{nbudaiev@ufl.edu}  

\author[0000-0001-6431-9633]{Adam Ginsburg}
\affiliation{Department of Astronomy, University of Florida, P.O. Box 112055, Gainesville, FL 32611, USA}
\email{adamginsburg@ufl.edu}

\author[0000-0003-0410-4504]{Ashley~T.~Barnes}
\affiliation{European Southern Observatory (ESO), Karl-Schwarzschild-Stra{\ss}e 2, 85748 Garching, Germany}
\email{ashley.barnes@eso.org}

\author[0000-0003-0416-4830]{Desmond Jeff}
\affiliation{Department of Astronomy, University of Florida, P.O. Box 112055, Gainesville, FL 32611, USA}
\affiliation{National Radio Astronomy Observatory (NRAO), 520 Edgemont Road, Charlottesville, VA 22903, USA}
\email{astrosynth1867@protonmail.com}

\author[0000-0003-2968-5333]{Taehwa Yoo}
\affiliation{Department of Astronomy, University of Florida, P.O. Box 112055, Gainesville, FL 32611, USA}
\email{t.yoo@ufl.edu}

\author[0000-0002-6073-9320]{Cara Battersby}
\affiliation{Department of Physics, University of Connecticut, 196A Auditorium Road, Unit 3046, Storrs, CT 06269, USA}
\email{cara.battersby@uconn.edu}

\author[0000-0002-4407-885X]{Alyssa Bulatek}
\affiliation{Department of Astronomy, University of Florida, P.O. Box 112055, Gainesville, FL 32611, USA}
\email{abulatek@ufl.edu}

\author[0000-0002-1313-429X]{Savannah Gramze}
\affiliation{Department of Astronomy, University of Florida, P.O. Box 112055, Gainesville, FL 32611, USA}
\email{savannahgramze@ufl.edu}

\author[0000-0003-2619-9305]{Xing Lu}
\affiliation{Shanghai Astronomical Observatory, Chinese Academy of Sciences, 80 Nandan Road, Shanghai 200030, P.\ R.\ China}
\affiliation{State Key Laboratory of Radio Astronomy and Technology, A20 Datun Road, Chaoyang District, Beijing, 100101, P.\ R.\ China}
\email{xinglv.nju@gmail.com}

\author[0000-0001-8782-1992]{Elisabeth A.C. Mills}
\affiliation{Department of Physics and Astronomy, University of Kansas, 1251 Wescoe Hall Drive, Lawrence, KS 66045, USA}
\email{eacmills@ku.edu}

\author[0009-0001-8880-6951]{Theo Richardson}
\affiliation{Department of Astronomy, University of Florida, P.O. Box 112055, Gainesville, FL 32611, USA}
\email{terichard57@gmail.com}

\author[0000-0001-7330-8856]{Daniel~L.~Walker}
\affiliation{UK ALMA Regional Centre Node, Jodrell Bank Centre for Astrophysics, The University of Manchester, Manchester M13 9PL, UK}
\email{daniel.walker.astro@gmail.com}

\begin{abstract}
We report JWST NIRCAM and MIRI observations of Sgr B2, \rr{one of} the most active site\rrr{s} of star formation in the Galaxy.
These observations, using 14 filters spanning 1.5 to 25 microns, have revealed a multilayered and highly structured cloud that contains both a revealed, low-extinction and hidden, high-extinction population of massive stars.
JWST has detected new candidate \hii regions around massive stars previously missed by radio telescopes.
MIRI has detected radiation escaping from the forming massive cluster Sgr B2 N along its outflow cavities, demonstrating that infrared radiation finds geometric escape routes even in the densest, most heavily embedded regions in the universe.
JWST further highlights the gas asymmetry in the cloud, showing a sharp, straight cutoff along the eastern cloud edge.

Despite the great sensitivity of these observations, no extended population of YSOs has been detected, placing a limit on their minimum extinction; this result hints that star formation has only just begun in the cloud.
Together, these results suggest that, despite already holding the crown for \rr{one of the} most actively star-forming cloud\rrr{s}, we have underestimated the total star formation in Sgr B2. JWST unveils previously hidden massive stars and ionized structures, offering a \rr{clearest-yet} view of how stars form under some of the most extreme Galactic conditions.
\end{abstract}

\section{Introduction}\label{sec:introduction}
\begin{figure*}
    \centering
    \includegraphics[width=1\linewidth]{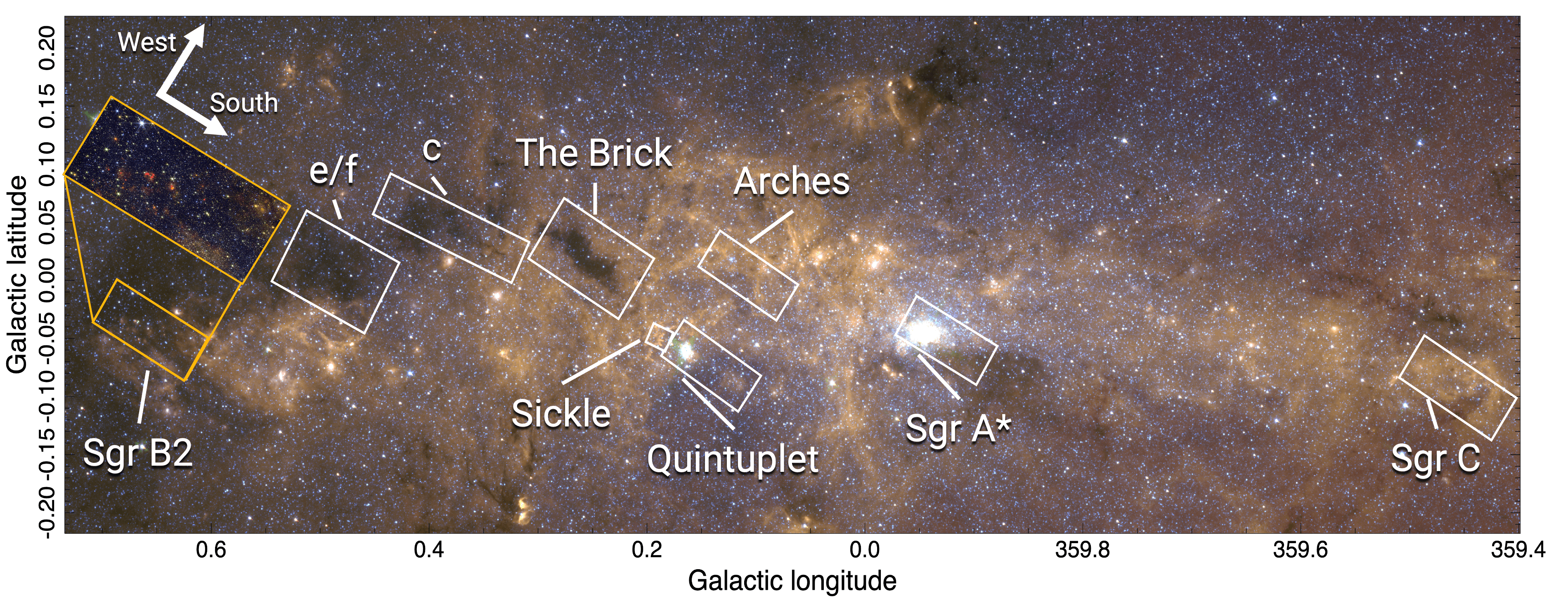}
    \caption{The overview of the CMZ with a \textit{Spitzer} IRAC tricolor image. The footprints of the currently completed JWST programs in the CMZ are shown with white rectangles. These include observations of molecular clouds Sgr B2 (this work, PID: 5365), cloud e/f (PID: 2092), cloud c (PID: 2221), The Brick (PIDs: 1182, 2221 \citep{ginsburg_jwst_2023}, 2526, 6927), and Sgr C (PID: 4147 \citep{crowe_jwst-nircam_2025}), star clusters Arches and Quintuplet (PID: 2045), as well as Sickle region (PID: 3958) and Sgr A*/nuclear star cluster (PIDs: 1939, 3571, 5368, 6095).
    The relative orientation of the top panel of Figure \ref{fig:NIRCam_and_MIRI} is shown with a yellow rectangle. 
    An interactive view of the JWST observations of the Galactic center can be viewed at: \href{https://starformation.astro.ufl.edu/avm_images/jwst_cmz_hips/}{https://starformation.astro.ufl.edu/avm\_images/jwst\_cmz\_hips}.}
    \label{fig:cmz_overview}
\end{figure*}

Stars form differently in star clusters, with probable differences in the initial mass function \citep[IMF, e.g.,][]{hosek_unusual_2019}, multiplicity \citep{duchene_is_2018}, and planets \citep{armitage_suppression_2000,daffern-powell_evaporation_2022}.
Such clusters represent a large fraction, in some cases even the majority, of star formation in the earliest galaxies \citep[e.g.][]{kruijssen_fraction_2012,kruijssen_comparing_2013,pfeffer_e-mosaics_2018}.
The Central Molecular Zone (CMZ\rr{, see Figure \ref{fig:cmz_overview}}) of our Galaxy is an excellent local laboratory in which to study massive cluster formation.
\rr{The environment in the CMZ is more extreme than that of the solar neighborhood: dust and gas temperatures, linewidths, magnetic field strengths, gas surface densities, and stellar densities are elevated by factors ranging from a few to several orders of magnitude \citep[see][and references therein]{henshaw_star_2022}. 
These conditions more closely resemble those prevalent during the most active period of cosmic star formation (z $\approx$ 2) \citep{madau_cosmic_2014, kruijssen_comparing_2013}.}
The extreme gas properties drive differences in the star formation process \rr{in the CMZ, such as having a higher fraction of stars form in dense, massive clusters }\citep{walker_comparing_2016, ginsburg_high_2018, Barnes2019}.

Despite containing around 80 percent of the Galaxy's dense molecular gas, the CMZ only forms around 10 percent of the Galaxy's stars, more than an order of magnitude lower from what we expect according to the typical dense gas relations \citep{barnes_star_2017,longmore_variations_2013}. 
Investigating and characterizing the star-forming population within the extreme environment of the CMZ is crucial to understand the low star formation rate. 

\rr{Sagittarius B2 (Sgr B2) is a powerful laboratory for studying star formation and evolution in extreme conditions compared to the solar neighborhood}.
The molecular cloud is situated $\sim$100\pc away in projection from the Galactic center, \rr{which is} located 8.277 \kpc away \citep{gravity_collaboration_mass_2022}. \rr{As a canonical member of the CMZ, Sgr B2 is generally assumed to lie at the Galactic center distance to within \rrr{a few hundred \pc.}}

Sgr B2 is forming stars at a rate of 0.04 \msun \yr$^{-1}$, almost half of all the star formation in the CMZ, and thus is one of the most star-forming clouds in our Galaxy \citep{ginsburg_distributed_2018}. 
Interferometry with long baselines has enabled resolved studies of the star-forming populations in the cloud;
the cloud contains over 700 YSOs, over 50 \hii regions, many outflows, dozens of hot cores, and hundreds of masers \citep{ginsburg_distributed_2018, budaiev_protostellar_2024, schmiedeke_physical_2016, meng_physical_2022, jeff_thermal_2024, budaiev_properties_2025}. JWST's sensitivity and resolution allows us to probe recent, but less embedded, star formation on similar scales.

We present the first findings of the JWST NIRCam and MIRI observations of Sgr B2. In Section \ref{sec:observations} we describe the observational setup and data reduction procedures. Section \ref{sec:analysis} explains the creation of an extinction map and SED fitting. 
In Section \ref{sec:discussion} we present new and unanticipated findings and discuss the implications with the conclusions summarized in Section \ref{sec:conclusions}.


\begin{deluxetable*}{cccc}
\tablecaption{Observation setup for JWST NIRCam and MIRI imaging.\label{tab:nircam_miri_obs}}
\tablehead{
\colhead{Filter\rr{\tablenotemark{a}}} &
\colhead{Bandwidth ($\Delta\lambda$, \microns)} &
\colhead{This Work's Primary Use} &
\colhead{Total Exposure Time (s)}
}
\startdata
F150W       & 0.318 & Continuum & 3092 \\
F182M        & 0.238 & Continuum for F187N & 3092 \\
F187N        & 0.024 & \paa recombination line & 12884 \\
F210M        & 0.205 & Continuum for F212N & 3092 \\
F212N        & 0.027 & \htwo & 3092 \\
F300M        & 0.318 & \water ice & 3092 \\
F360M        & 0.372 & Continuum & 3092 \\
F405N+F444W  & 0.046 & \bra recombination line & 9792 \\
F410M        & 0.436 & Continuum for F405N & 3092 \\
F466N+F444W  & 0.054 & CO ice & 3092 \\
F480M        & 0.303 & Continuum & 3092 \\
\hline
F770W        & 1.95   & PAH, hot dust & 236 \\
F1280W       & 2.47   & Hot dust, continuum & 236 \\
F2550W       & 3.67   & Warm dust, continuum & 236 \\
\enddata
\rr{\tablenotetext{a}{The filter naming scheme is representative of the filter's effective wavelength in microns$\times$100 to within $\sim$0.05\microns.}}
\end{deluxetable*}


\section{Observations and data reduction}\label{sec:observations}
JWST observed the extended Sgr B2 cloud as a part of program 5365 (PIs: A. Ginsburg, N. Budaiev). 
The data presented in this paper were obtained from the Mikulski Archive for Space Telescopes (MAST) at the Space Telescope Science Institute (STScI). 
The data described here may be obtained from the MAST archive at
\dataset[doi:10.17909/T9RP4V]{https://dx.doi.org/10.17909/p90j-rw02}.
NIRCam observations were completed on September 7th 2024. 
MIRI observations were split into two visits: the northern part of the cloud was observed on September 15th 2024, and the remaining half was observed on September 2nd 2025.

The observations include 11 NIRCam filters and 3 MIRI filters. The majority of the NIRCam filters used the SHALLOW2 readout pattern with three groups per integration to reduce saturation in the bright stars; the filter pair with the longest exposure, F187N and F405N, used the MEDIUM8 readout pattern to reduce the data rate. \rr{The F187N filter, covering \rrr{the} \paa recombination line was included in two filter pair observations to maximize the sensitivity.}
\rr{All NIRCam data used 1 integration per exposure, 24 dither positions, each with 24 integrations. The MIRI observations all used the FASTR1 readout, with 8 groups per integration, 2 integrations per exposure, 1 exposure per dither, 5 dither positions using the ``cycling large" dither pattern, with 10 integrations. The total exposure times for each filter as well as the primary uses are given in Table \ref{tab:nircam_miri_obs}.}
The observations were ordered by increasing filter bandwidth and from shortest to longest wavelength to mitigate persistence. 
The MIRI observations did not include a background pointing. Due to broad and bright emission throughout the CMZ, there is not a suitable location to measure true background. 

\begin{figure*}
    \centering
    \includegraphics[width=0.99\linewidth]{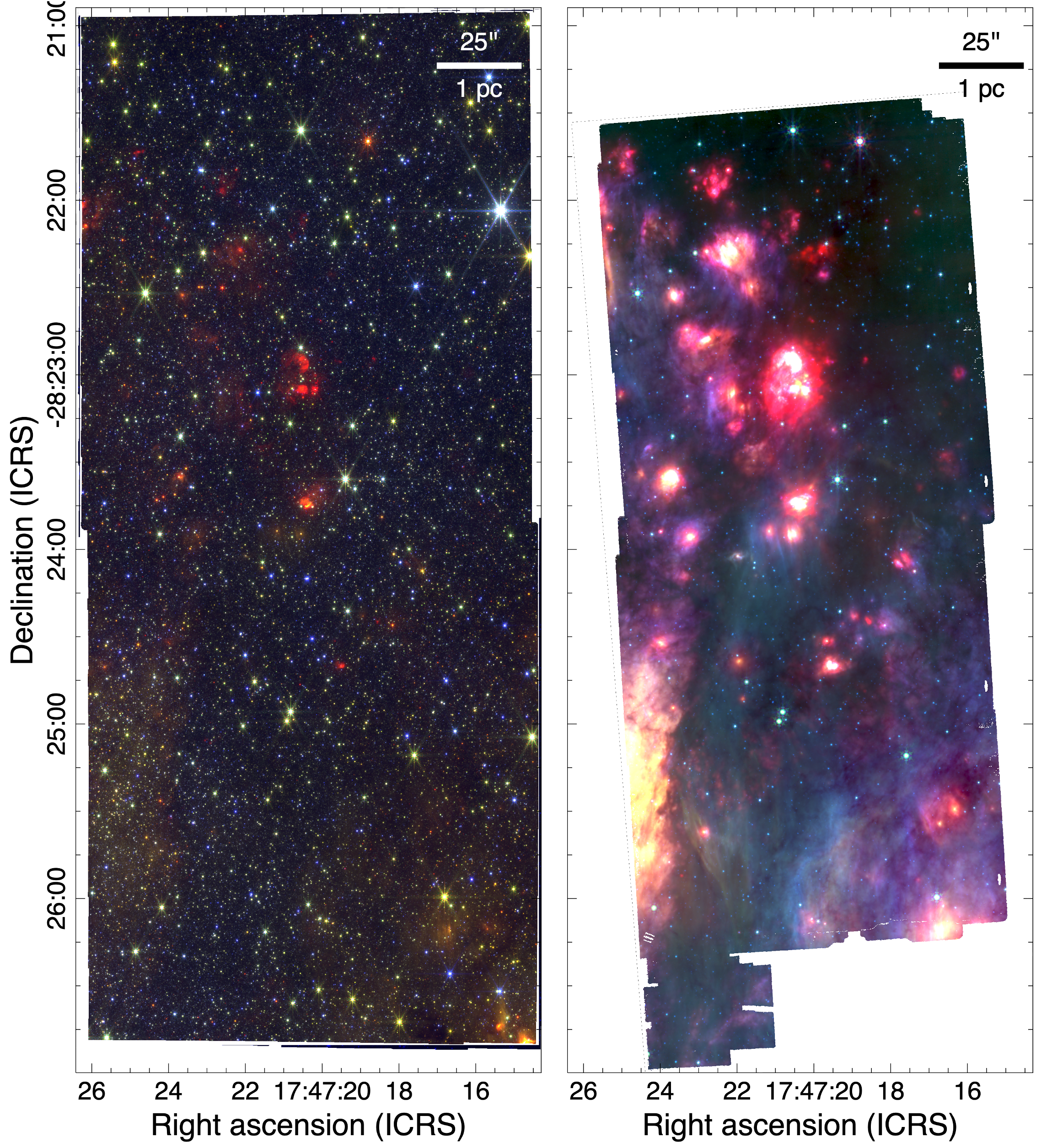}
    \caption{\rrr{Left}: NIRCam mosaic of Sgr B2 with F480M in red, F360M in green, and F150W in blue. 
    \rrr{Right}: MIRI mosaic of Sgr B2 with F2550W in red, F1280W in green, and F770W in blue. \rr{The presented filters trace primarily continuum emission, except F770W, which traces PAH emission.}\footnote{\rr{See also NASA's press release for an alternative rendering of NIRCam and MIRI mosaics, including an interactive slider: \href{https://science.nasa.gov/missions/webb/nasas-webb-explores-largest-star-forming-cloud-in-milky-way/}{https://science.nasa.gov/missions/webb/nasas-webb-explores-largest-star-forming-cloud-in-milky-way/}}}}
    \label{fig:NIRCam_and_MIRI}
\end{figure*}




\subsection{Imaging}
We imaged the data using STScI's science calibration pipeline version 1.15.1.
The NIRCam imaging was done with \texttt{suppress\_one\_group=False} to recover some of the saturated areas. An overview tricolor image is shown in the \rrr{left} panel of Figure \ref{fig:NIRCam_and_MIRI}.
The MIRI data, especially in the F2550W filter, are saturated around the bright \hii regions.
Thus, MIRI data were imaged twice: once for source cataloging and once for image presentation. The version for the source cataloging used the standard pipeline parameters with an addition of sky subtraction with \texttt{skymatch} set to \texttt{match}. 
We imaged the data again with all saturation flagging turned off.
These latter images are used purely for better visual presentation of the figures in this work.
The MIRI tricolor image is presented in \rrr{right} panel of Figure \ref{fig:NIRCam_and_MIRI}.
\begin{figure*}
    \centering
    \includegraphics[width=0.8\linewidth]{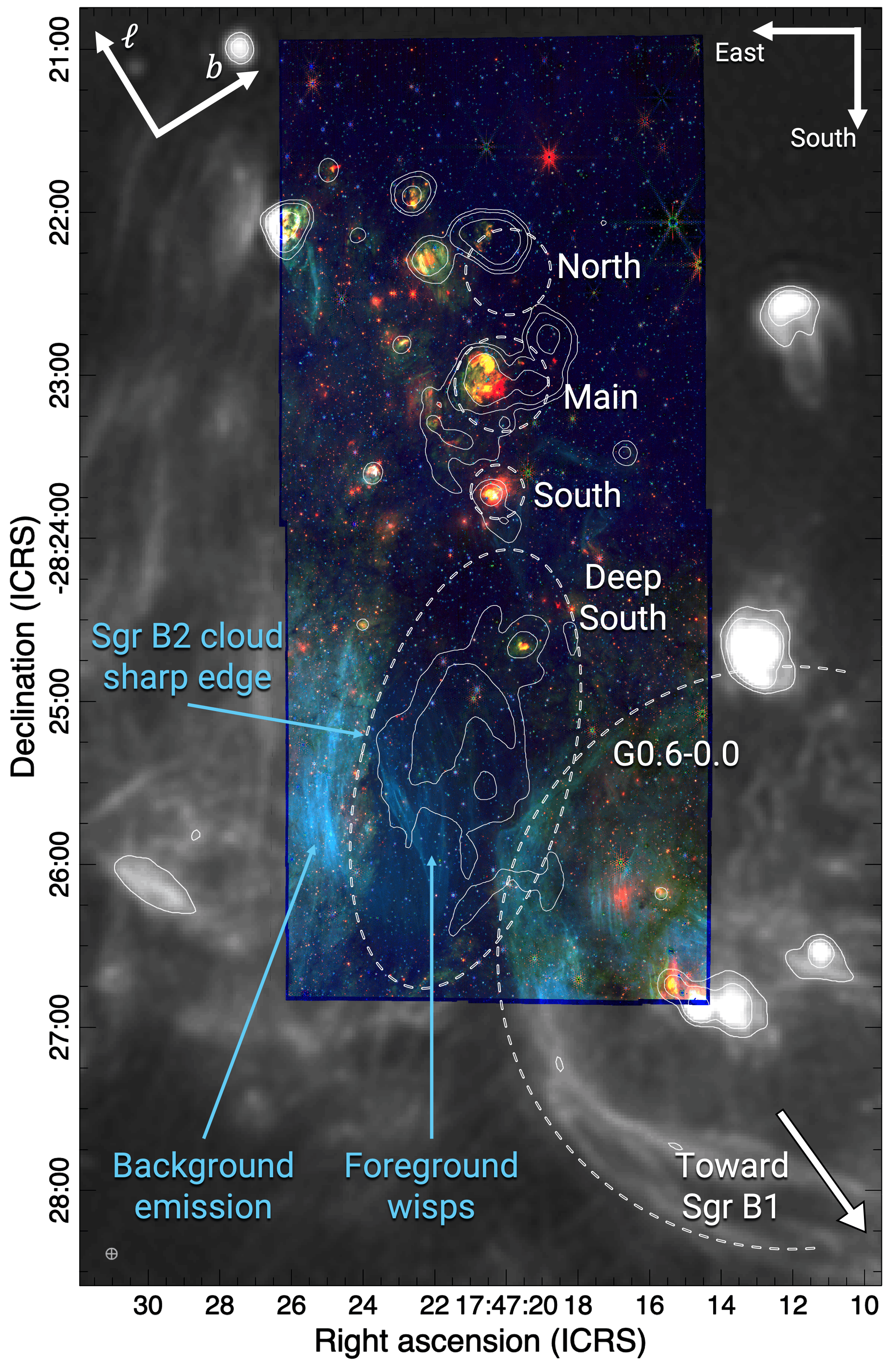}
    \caption{A NIRCam tricolor figure overlaid on top of MeerKAT 1.2\ghz continuum with the main parts of the cloud and the observed features labeled. \rrr{Red is 4.8 micron excess (F480M-[F410M-F405N]) green is continuum-subtracted \bra ([F405N-F410M]), and blue is continuum-subtracted \paa ([F187N-F182M]).} The blue color represents low extinction, green shows regions with higher extinction, and red shows warm dust. \rr{The East (left) side of the cloud contains recently formed stars seen with JWST, while the West (right) side of the cloud contains a lot of dusty ALMA-detected YSOs and is much less populated in the JWST filters.}}
    \label{fig:overview_labels}
\end{figure*}
Figure \ref{fig:overview_labels} shows the cardinal directions relative to the JWST footprint as well as the main features of the cloud. \rf{The four distinct regions of the cloud are labeled: two massive protoclusters Sgr B2 N(orth) and Sgr B2 M(ain), a less massive and younger protocluster Sgr B2 S(outh), and an extended star-forming dust ridge Sgr B2 D(eep) S(outh).}

\subsection{Source extraction}\label{sec:source_extraction}
The source extraction for NIRCam filters is performed following the methods described in \cite{ginsburg_jwst_2023} (see their Sections 3.3 and 3.4). In summary, \texttt{crowdsource} in combination with \texttt{webbpsf} and \texttt{stpsf} \citep{perrin_simulating_2012,perrin_updated_2014} are used on the individual calibrated exposures to produce a catalog for each frame, which are subsequently merged. 
The catalog\rrr{s} for each filter are then combined by cross-matching detections and excluding any matches with a separation of $d>0.\arcsec1$. 
We then filter the catalog by ``quality factor" \texttt{qf} $>0.75$, how extended the source is \texttt{spread} $< 0.25$, magnitude error \texttt{emag} $<0.1$, and fraction of flux attributed to the source's PSF \texttt{fracflux} $>0.8$.
There are 1,293,208 sources with a good measurement in at least one of the filters, 
84,490 sources with a good measurement in all of the filters where the source was detected,
and 29,006 sources with a good measurement in all eleven filters.

We utilize the STScI's pipeline implementation of \texttt{DAOphot} to extract sources in the three MIRI filters. We set the \texttt{kernel\_fwhm} parameter to the PSF's FWHM in pixels for each filter.
Since observing a dedicated background field is not feasible for a Galactic center target, the true background level is uncertain. The imaging utilizes the \texttt{skymatch} method set to \texttt{match}, matching down. This strategy allows for smooth background matching between the frames without making assumptions about the large-scale background. The background-subtracted aperture photometry performed on point-sources is sufficient to exclude any large-scale background contribution. 

The MIRI images contain a lot of smaller-scale extended but structured emission. We choose \texttt{bkg\_boxsize} of 60 pixels for F2550W, 35 pixels for F1280W, and 25 pixels F770W to improve the source extraction in the regions close to the bright extended emission. 
We then exclude any detection with \texttt{is\_extended} flag, resulting in 3726 detections at 7.7\microns, 1612 detections at 12.8\microns, and 209 detections at 25.5\microns.
We note that, while the \texttt{daophot}'s \texttt{is\_extended} classification performs well at the shorter wavelengths, more than half of sources classified as point sources in F2550W are instead extended. This brings down the number of true point sources to about 50, not including some of the faint objects missed by the cataloging tool. Finally, the F2550W filter observations are affected by persistence from the preceding observations with F1280W filter (see Appendix \ref{sec:persistence}).

\rr{Source extraction tools (e.g. \texttt{crowdsource}, \texttt{daophot}) struggle with extremely crowded fields towards the Galactic center. The images are further complicated by the bright extended emission and several extremely bright stars (H mag $<9$) and thus extended PSF artifacts.} 
After performing by-eye inspection and preliminary analysis \rrr{of our NIRCam and MIRI source catalogs}, it is evident that, while these catalogs are sufficient to identify the general features of the observed populations, a more sophisticated approach is needed to isolate the \rr{rarer} populations (e.g. YSOs, UC\hii regions).
We subjectively rank the main factors impacting the catalog uncertainty, from most to least important: source density, extended psf artifacts, and extended emission for NIRCam. MIRI catalogs are primarily affected by the structured extended emission.
An in-depth performance analysis of cataloging tools as well as source extraction uncertainty mitigation will be performed elsewhere.

\subsection{Other data}
In this work, we compare the JWST data to ALMA \rr{observations with program IDs 2016.1.00550.S \citep[1 and 3 mm continua;][]{budaiev_protostellar_2024}, 2013.1.00269.S \citep[3 mm continuum;][]{ginsburg_distributed_2018}, and 2017.1.00114.S (1 mm continuum and SiO J=2$–$1 line), MeerKAT data \citep[1.2\ghz continuum;][]{heywood_128_2022}, and VLA data \citep[6\ghz continuum;][]{meng_physical_2019}, \citep[\rrr{22\ghz continuum;}][]{Gaume1995},  \citep[22\ghz \water maser line and 22\ghz continuum;][]{budaiev_properties_2025}.}
\rf{The far-infrared Herschel observations were not utilized in the multi-wavelength analyses because the \pc-scale resolution of these data are not sufficient to substantiate or refute the conclusions presented in this work.}

\section{Analysis}\label{sec:analysis}


\subsection{Recombination line extinction map}\label{sec:extinction_map}
Extinction is wavelength dependent; thus, the intrinsic ratio of the emissivities of recombination lines can be used to derive the extinction. 
\rr{For an optically thin medium e}missivity is defined as
\begin{equation}
    j_\nu \equiv \frac{d I_\nu}{d s},
\end{equation}
where $I_\nu$ is the surface brightness\rr{, and $s$ is the path length along the line of sight}. If the emission is optically thin, then the path length of \paa and \bra is the same, and thus
\begin{equation}
    \frac{I_{\paa}}{I_{\bra}} = \frac{j_{\paa}}{j_{\bra}}.
\end{equation}
The pipeline-delivered data are in \Mjysr, so in order to compare the line surface brightness in different filters, we need to account for the bandwidth difference. Assuming no extinction, the ratios of observed bandwidth-scaled lines and emissivities should be the same:
\begin{equation}
    \frac{I_{F187N} \times BW_{F187N}}{I_{F405N} \times BW_{F405N}} = \frac{j_{\paa}}{j_{\bra}} = R_0,
\end{equation}
where $I_{filter}$ is the observed surface brightness in the given filter, and $BW_{filter}$ is the bandwidth of the filter.
At T=10$^4$ K, $\frac{j_{\paa}}{j_{\bra}} = 4.24$ \citep{draine_physics_2011}. At T=$5\times10^3$ K, the ratio drops to 3.96.

Let $k_{\lambda} = \frac{A_{\lambda}}{A_V}$, where $k$ is the assumed extinction curve.
Then the extinction is:
\begin{equation}
    A_V=\frac{2.5 \log _{10}\left(R_0 / R_{obs}\right)}{\left(k_{\lambda_1}-k_{\lambda_2}\right)},
\end{equation}
where $R_{obs}$ is the observed ratio, $R_0$ is the intrinsic ratio.

We produce continuum-subtracted recombination line images by first subtracting the line contribution from the corresponding continuum filter image and then subtracting the resulting line-free continuum image from the narrow-band data. We refer the reader to the Section 3.5 of \cite{ginsburg_jwst_2023} for the detailed explanation of the continuum-subtracted line image creation. 

\begin{figure}
    \centering
    \includegraphics[width=\columnwidth]{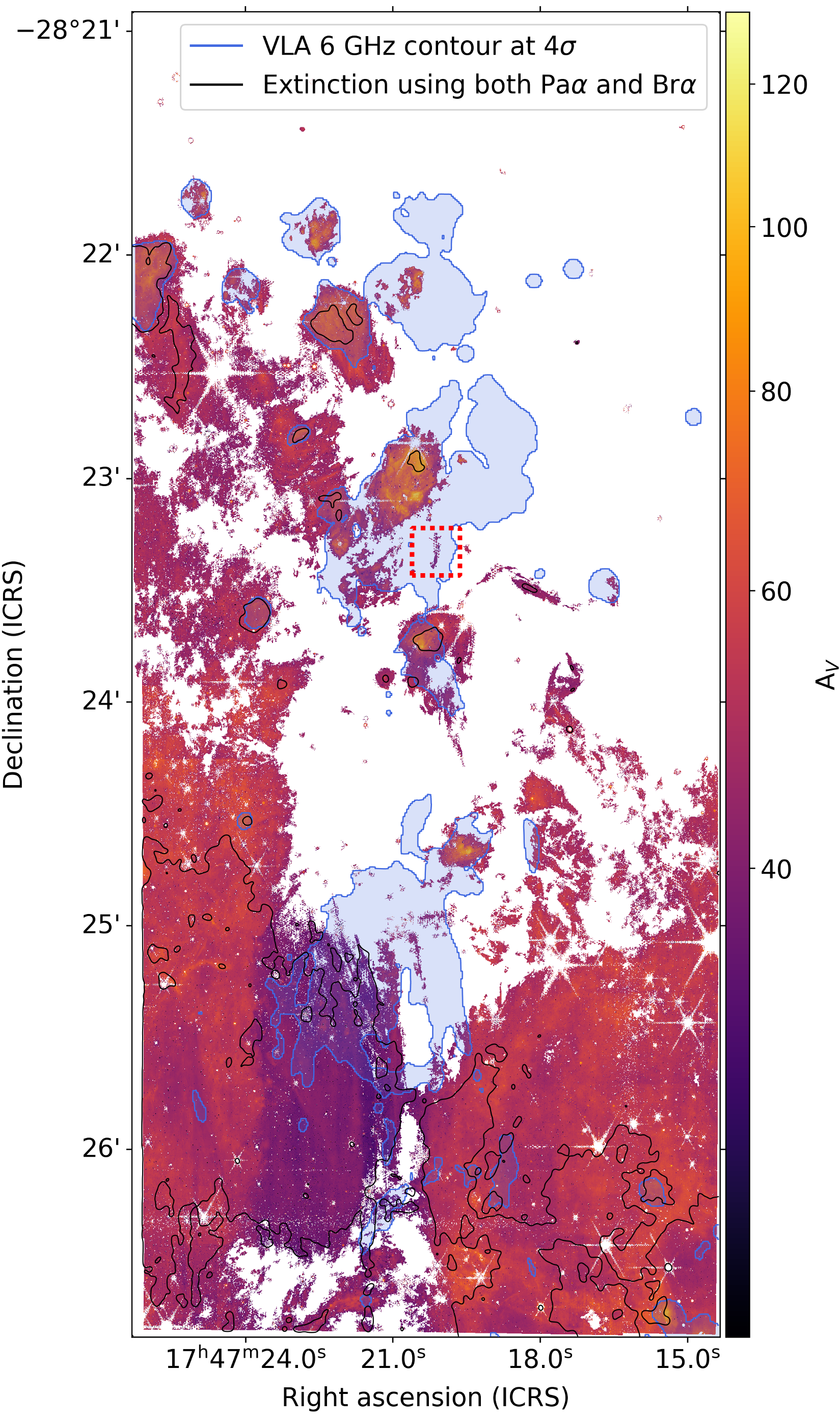}
    \caption{Extinction map based on the \paa and \bra line ratio. The black contours define the areas where both \bra and \paa emission is present. Outside of the black contours, only \bra emission is detected, setting a lower limit on extinction. The apparent dark patch of the map on the \rr{bottom} of the image is coincident with Sgr B2 DS, a region with high column density. The measured extinction is towards a foreground feature; the Sgr B2 DS region is extremely deeply embedded such that ionized material behind it is it not detected in either Pa$\alpha$ or Br$\alpha$. 
    \rr{6 GHz VLA emission from \cite{meng_physical_2019} \rf{shaded} in blue shows the extent of free-free emission.}
    \rf{The blue-on-white shaded regions can be interpreted as having A$_V \gtrsim 100$, while the white areas do not provide any information on extinction due to lack of recombination line emission.}
    The red dotted square marks the location of one of the few features with an \htwo detection.}
    \label{fig:extinction_bra}
\end{figure}

We assume an extinction law from \cite{chiar_pixie_2006}\footnote{\texttt{CT06-MWGC} extinction curve in \texttt{dust\_extinction} package.} to compute the $A_V$ over the whole FOV. 
This extinction law is chosen as it is suitable for the Galactic center and covers all of the observed filters.
Then, we visually inspect the \paa and \bra images to identify the minimum flux where the extended emission can be distinguished from the background 1/f noise: 1.5\Mjysr for \paa and 3\Mjysr for \bra. \rr{After convolving \paa image to the lower resolution of the \bra image, w}e create two maps using these emission thresholds as masks: the \paa -based map is constrained to regions that have emission in both lines, while the \bra -based map includes regions that do not have \paa emission present, but do have the \bra emission visible. The regions with only \bra emission are still useful as they set a lower limit on the extinction.

The resulting images contain visual artifacts from the bright stars; the short- and medium-band filters have significantly different PSFs and thus the continuum-subtracted images contain characteristic spikes. The different size of the PSF prevents a proper subtraction of faint stars, as well, resulting in small, isolated groups of pixels with non-physical extinction values. 
We remove the diffraction artifacts and the isolated stars from the extinction maps by removing isolated groups of pixels via \texttt{remove\_small\_objects} from \texttt{scipy.morphology}. We set the minimum size at 30 and 250 pixels for \paa- and \bra-based maps respectively. In addition, 1/f noise in the shorter wavelength images results in faint \rr{vertical} striping in the resulting extinction map, primarily on the \rr{lower} side of the image.

\begin{figure*}
    \centering
    \includegraphics[width=1\linewidth]{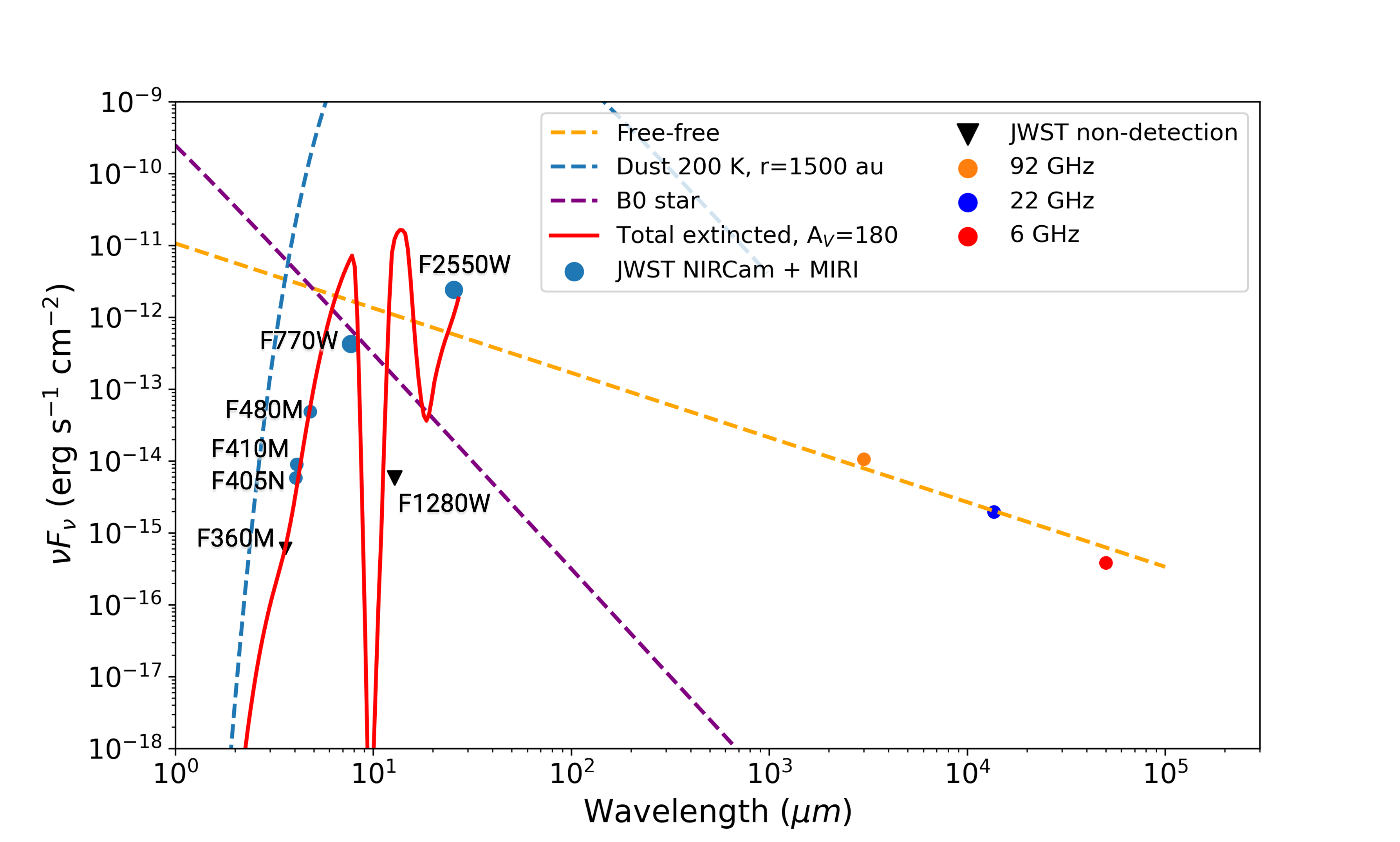}
    \caption{An SED of the X8.33 UC\hii region. 
    The contribution of the free-free emission is set by the observations in the radio regime. 
    The stellar type of the central star, and thus the stellar contribution, is derived from the estimates of the number of ionizing photons from \cite{meng_physical_2022}. The dust contribution is modeled as a blackbody based on the observed size in the 3 mm observations. The remaining free parameters $T_{d}$ and $A_V$ are fit to match the observed SED. The JWST filter bandwidths are smaller than the marker sizes. The flux error bars are similar to the marker sizes.} 
    \label{fig:X833}
\end{figure*}

The \bra-based extinction map is shown in Figure \ref{fig:extinction_bra}. The black contours enclose the regions where both \paa and \bra recombination lines are detected; any point outside the contours is only detected in \bra and thus is a lower limit on the extinction. In such cases, we still use the pixel values from the \paa map instead of a constant value to compute the lower limit on extinction. This choice results in a visually smoother map.
This measurement will be discussed further in Section \ref{sec:extinction}.

\subsection{SED modeling methodology}\label{sec:SED}
\rr{We construct an SED of UC\hii region X8.33 which is detected with VLA, ALMA, and now JWST. This \hii region is isolated, resolved in ALMA data, and detected in the greatest number of JWST filters, which simplifies the interpretation of the SED compared to some of the more complex sources.}
\rr{To determine the emission mechanisms contributing to the JWST flux for this HII region, we adopt several single-component models as limiting cases.  We evaluate pure stellar, pure dust, and pure ionized gas to establish that no single-component model fits. Instead, we construct a rudimentary model that combines free-free, dust, and stellar contributions. The free-free emission is constrained based on millimeter flux. 
The stellar contribution, which constrains extinction using shorter-wavelength data, is determined from previously measured rates of ionizing photons in radio observations. Finally, the MIRI filters show that a dust component is required.}
\rr{Fig \ref{fig:X833} shows the HII region's SED being fit as a combination of free-free emission, warm dust, and stellar emission behind a screen of $A_V\gtrsim180$.}


The X8.33 \hii region is located west of Sgr B2 N, next to ALMA-detected YSOs (see Figure \ref{fig:HII_cand_asymmetry}). The long-baseline 3 mm data spatially resolves the source to have $r = 1500$ au (0.014\pc). Using the 6, 22, and 96 \ghz data \citep{meng_physical_2022, Gaume1995, budaiev_protostellar_2024}, we determine the spectral index in the radio regime: $\sim$0.15 between 92 and 22\ghz and $\sim$0.25 between 22 and 6\ghz. 
The low spectral index suggests that the \hii region is dominated by free-free emission and is optically thin at 22\ghz.
\rr{We extrapolate the free-free emission based on the 22\ghz flux as this frequency is less affected by dust compared to 92\ghz and is less susceptible to optical depth effects than 6\ghz. However, extrapolating the free-free emission from a different frequency for this particular source does not significantly impact the conclusions.}
\rr{We model the free-free emission as $S_\nu \propto \nu^{-0.1}$. This relation should} hold until $h\nu \gtrsim kT_e$, which, assuming $T_e \sim 10^4$ K, should be valid until $\sim$1.5\microns \citep{rybicki_radiative_1979}. 


\rr{We then add the stellar contribution of a B0 central star based on the measured rate of ionizing photons $\log _{10}\left(\dot{N}_{\mathrm{Ly}} / \mathrm{s}^{-1}\right) \approx 46.6$ by \cite{meng_physical_2022}. We incorporate stellar emission in the SED using a blackbody model with $T=30,000$ and $r=7.4\rsun$ \citep{Pecaut2013}.}

\rr{We place the absolute lower limit on the extinction towards X8.33 by assuming no dust contribution to the emission. Using the \cite{chiar_pixie_2006} extinction law, we determine that a minimum of $A_V\gtrsim 140$ is required to explain the observed flux at 4\microns. }
\rr{Any additional dust contribution would increase the extinction.}
\rr{Indeed,} the significant excess emission observed \rf{at wavelengths greater than} 4.8\microns indicates the presence of hot dust that dominates at mid-infrared wavelengths. 
At 4\microns, blackbody radiation from the dust dominates at $T\ge190$ K assuming that the emitting dust has the same physical extent as the free-free emission seen at 92\ghz or $r= 1500$ au. An even higher temperature is required to explain the high flux observed in the F2550W filter. 
Figure \ref{fig:X833} shows that the observed SED is somewhat consistent with $T_d = 200$ K, $A_V = 180$ assuming an extinction curve from \cite{chiar_pixie_2006}. \rrr{This result is consistent with the extinction estimate of $A_V\sim200$ based on surface density reported in \cite{ginsburg_distributed_2018}.}
However, due to the uncertainty on the true shape of the extinction curve \citep{nogueras-lara_variability_2019} and a lack of direct temperature measurements, these are likely only rough estimates. 
\rrr{For instance,} \rr{using an extinction law from \citealt{gordon_one_2023} with $R_V=3.1$ results in an increase in the inferred $A_V$ by $\sim$30\% \citep{Decleir2022, Gordon2021, Fitzpatrick2019, Gordon2009}.}

\section{Discussion}\label{sec:discussion}
\subsection{Extinction}\label{sec:extinction}
In Section \ref{sec:extinction_map}, we created an extinction map of Sgr B2 using the recombination line ratio. 
The absence of measurements in the \rf{JWST-based} extinction map impl\rrr{ies} either an extinction $A_V\gtrsim$ \rf{100} or that no extended recombination line emission is present. 
\rf{The blue shaded regions in Figure \ref{fig:extinction_bra} show the extent of the free-free emission traced with 6\ghz VLA observations \citep{meng_physical_2019}. Thus, the blue-on-white shaded regions can be interpreted as $A_V\gtrsim$ 100, while the purely white regions do not provide any information on extinction. Further interpretations on the locations of high extinction regions can be done by identifying compact dust emission (see locations of ALMA-detected YSOs in Figure \ref{fig:HII_cand_asymmetry}).}
By comparing the extinction map with the existing mm and cm maps \citep[e.g.][]{ginsburg_distributed_2018, heywood_128_2022, meng_physical_2022}, we find that the majority of the cloud seen with ALMA and VLA is extremely embedded behind a thick screen of $A_V\gtrsim$ \rf{100}. 

\rf{We caution the reader that the JWST-measured extinction does not necessarily trace the cm-detected free-free emission.}. For example, \cite{meng_physical_2019} reports extended free-free emission in Sgr B2 DS and the corresponding region in the extinction map in Figure \ref{fig:extinction_bra} appears to have very low extinction. Based on the morphology of the features in the two data sets, they are likely to be distinct, with the JWST-observed wisps existing in the foreground of Sgr B2 DS.

The rapid change in the observed extinction in the \rr{lower} left part of the map suggests that the background extended emission located at the very \rrr{edge} of the image becomes completely \rr{extinguished} by the dense cloud. As mentioned above, the apparent lower extinction is the product of foreground emission on top of the star-forming ridge Sgr B2 DS. This feature is discussed further in Section \ref{sec:sharp_edge}.

Only a small fraction of ALMA- and VLA-detected parts of the cloud have an extinction measurement in JWST data. 
Sgr B2 S is seen at $A_V \approx 90$, and several \hii regions associated with Sgr B2 M are observed at $A_V \approx 100$. However, the bulk of the Sgr B2 M protocluster is completely \rr{extinguished} in both recombination lines, as is Sgr B2 N. 
The measured extinction to S and M \rr{could indicate} that these protoclusters began removing the surrounding dust through feedback, whereas Sgr B2 N is still deeply embedded. 
Typically, \rr{however}, due to the large number of \hii regions, Sgr B2 M is considered the oldest star forming region in the cloud, followed by Sgr B2 N, and then Sgr B2 S \citep{Qin2011, ginsburg_distributed_2018, meng_physical_2022}. \rr{Therefore,} an alternative interpretation for the observed extinction is that Sgr B2 S is located closer to the ``surface" of the cloud compared to Sgr B2 N.

Dense star forming regions contain a large amount of shocked gas, 
either from protostellar outflows colliding with the surrounding interstellar medium, or at sites of the cloud-cloud collisions.
Many outflows are detected in Sgr B2 DS based on SiO emission \citep{meng_physical_2022}, a known shock tracer. Another shock tracer covered with the NIRCam observations is \htwo.
\rr{We see no \htwo emission associated with the cloud \rf{, which we attribute to the high extinction.}}
\rf{The one exception is a} linear feature between Sgr B2 S and M, marked with a red dotted square in Figure \ref{fig:extinction_bra}. A part of this feature is seen in \bra with a lower limit $A_V = 40$. 


\begin{figure*}[h!]
    \centering

    \begin{subfigure}[t]{0.49\textwidth}
        \centering
        \includegraphics[width=\linewidth]{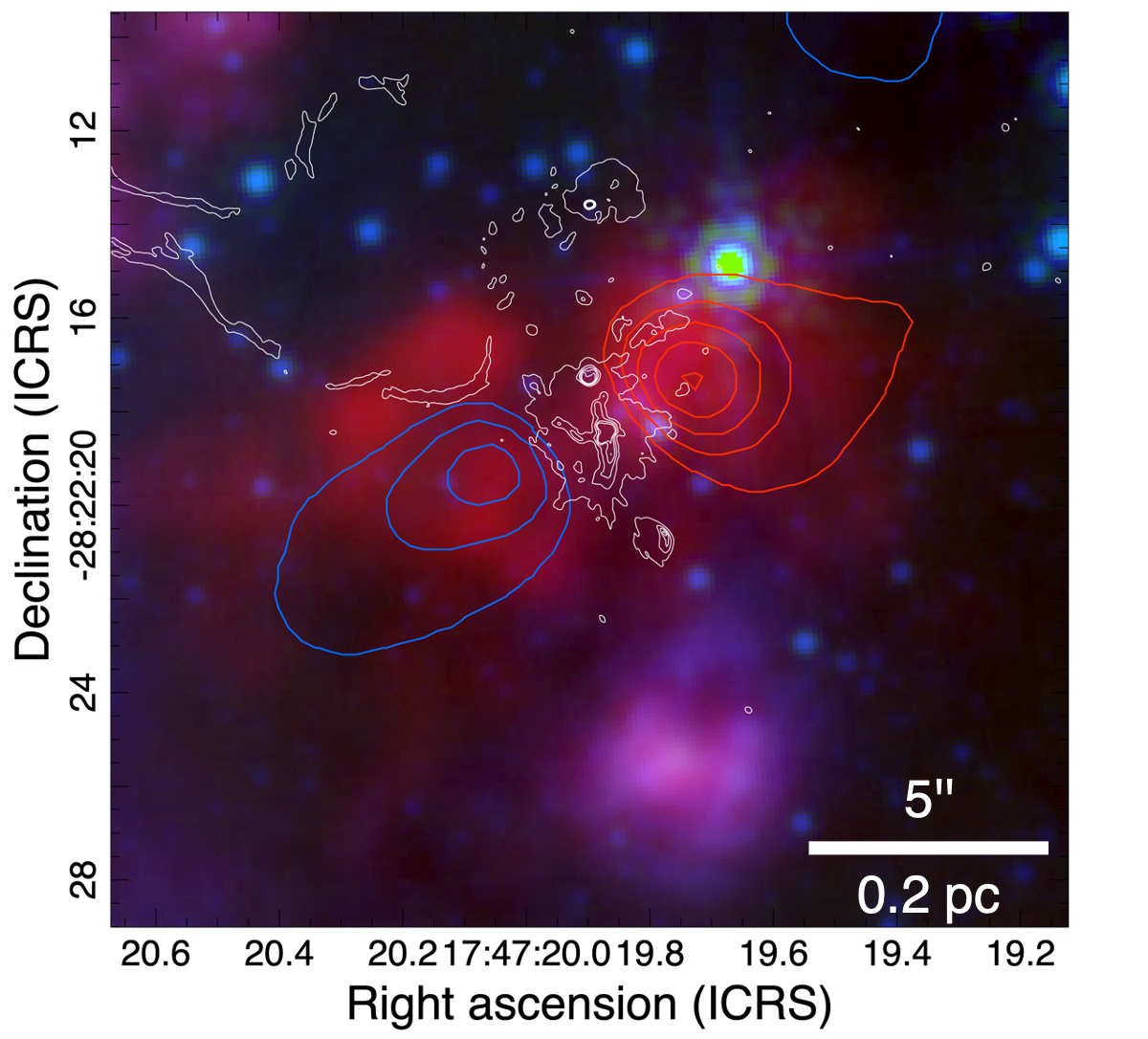}
        \caption{MIRI tricolor image: red is F2500W, green is F1280W, and blue is F770W. The 3 mm continuum emission from panel (b) is shown as white contours. The red and blue contours show the integrated intensity map of SiO emission between 5 and 55\kms and 85 and 115\kms as seen in ACES (S. Longmore et al. submitted, X. Lu et al. submitted). The 25\microns emission follows the extent and orientation of the outflow.}
    \end{subfigure}
    \hfill
    \begin{subfigure}[t]{0.49\textwidth}
        \centering
        \includegraphics[width=\linewidth]{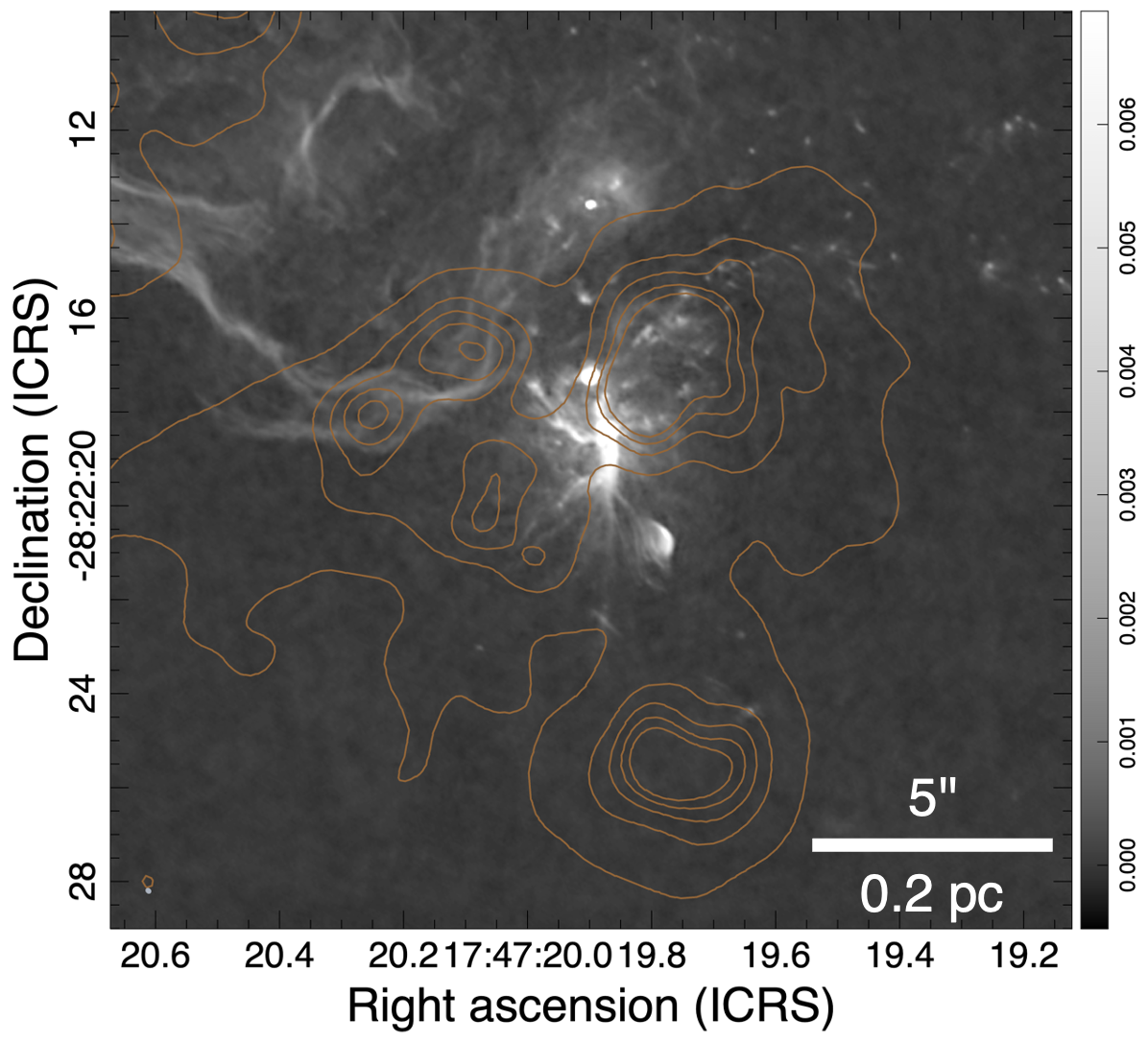}
        \caption{3 mm continuum ALMA from \cite{budaiev_protostellar_2024} that was imaged with \rr{shorter} baseline data from \cite{ginsburg_distributed_2018} to recover large angular scales.}
    \end{subfigure}


    \begin{subfigure}[t]{0.49\textwidth}
        \centering
        \includegraphics[width=\linewidth]{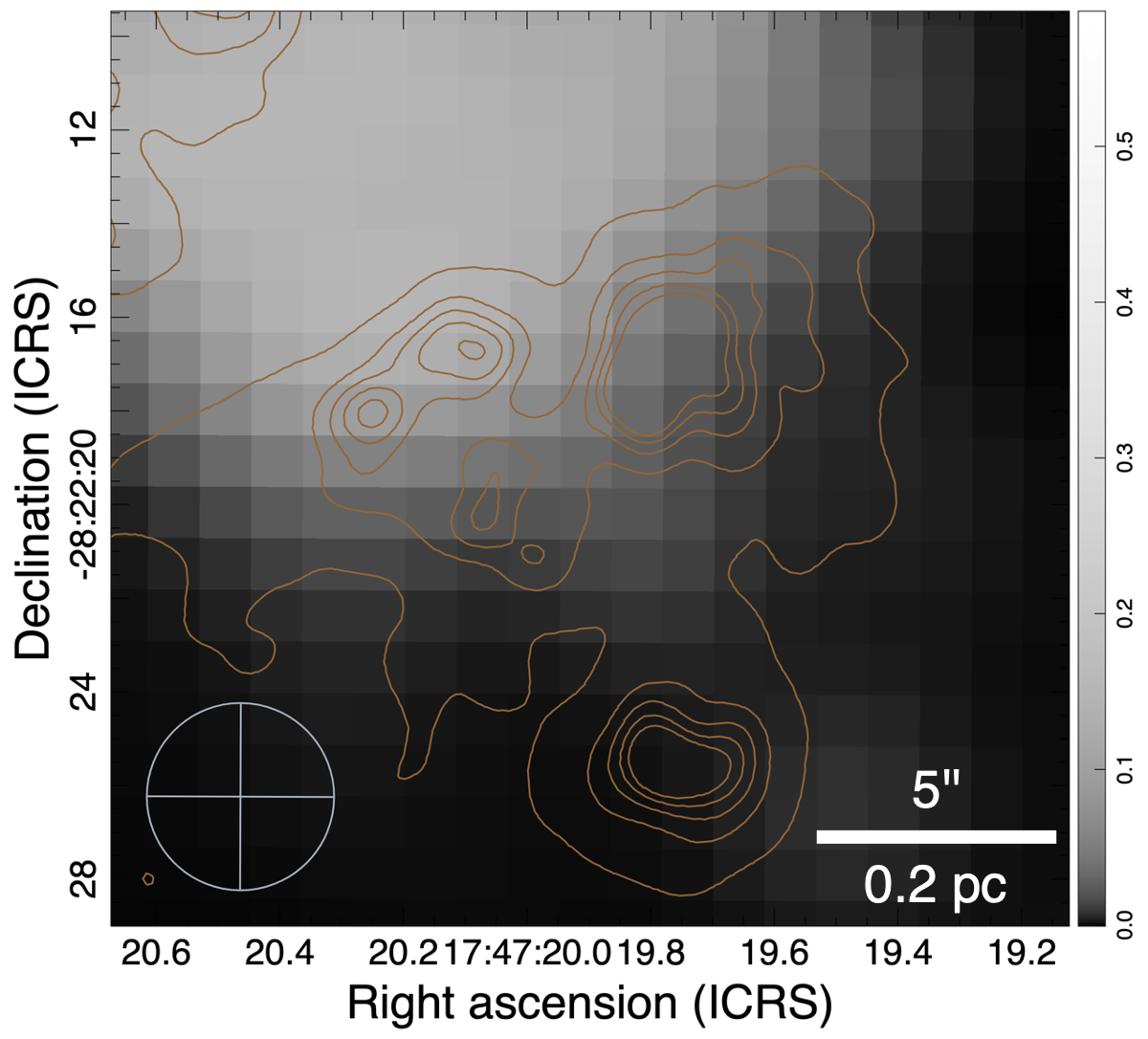}
        \caption{1.2 GHz continuum MeerKAT \citep{heywood_128_2022}.}
    \end{subfigure}
    \hfill
    \begin{subfigure}[t]{0.49\textwidth}
        \centering
        \includegraphics[width=\linewidth]{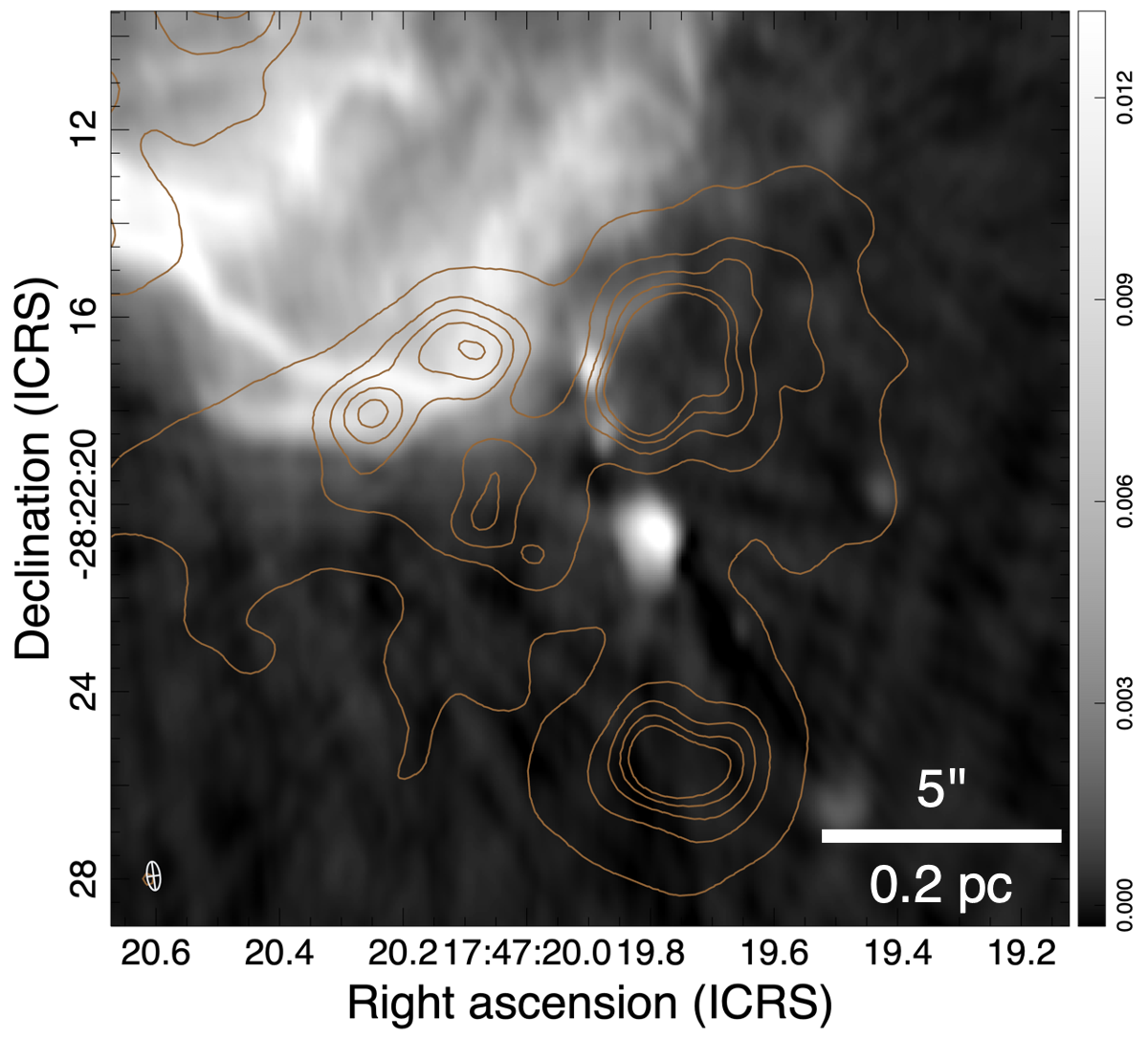}
        \caption{\rrr{6} GHz continuum VLA \citep{meng_physical_2022}.}
    \end{subfigure}

    \caption{A multiwavelength overview of the Sgr B2 N region. Panel (a) shows extended emission at 25\microns escaping the dense protocluster. Some of this emission was detected with \textit{Spitzer}, however due to the resolution and the presence of nearby \hii regions, seen as purple in this image, the nature of emission was ambiguous. The \rr{orange} contours show the MIRI 25.5\microns emission from panel (a). The beam sizes of the radio observations are shown as white ellipses on the bottom left.}
    \label{fig:SgrB2_N_multipanel}
\end{figure*}

\subsection{MIRI reveals radiation escaping Sgr B2 N}\label{sec:MIRI_SgrB2_N}
\rrr{There is an ongoing debate on the effect of the escaping radiation on the feedback in protoclusters. \cite{murray_disruption_2010} suggests that even despite the presence of low density ``holes", the radiation pressure is reduced by no more than 30\%. On the other hand, recent simulations by \cite{menon_infrared_2022, menon_outflows_2023} show that the radiation ends up escaping protoclusters and does not regulate star formation.}

As seen in Figure \ref{fig:SgrB2_N_multipanel}, there is extended 25\microns emission present surrounding the heart of the Sgr B2 N cluster. \rr{Within our subset of filters, t}his extended emission is detected only at 25\microns. \rr{There are no} radio counterparts between 1\ghz and 230\ghz. However, it matches the extent and orientation of the large-scale SiO outflow \citep{higuchi_sgr_2015, budaiev_properties_2025}. 
\rrr{This outflow was thought to be a group of smaller outflows collimated by the forming protocluster \citep{peters2012} as evidenced by its low velocity \citep[$\sim$15\kms;][]{higuchi_sgr_2015}. However, recent high-resolution radio observations show that the outflow likely originates from a single source and it is the outflow that sets the kinematics in the protocluster \citep{budaiev_properties_2025}.}

The observed configuration \rrr{in Sgr B2 N} suggests that infrared radiation escaping from the heart of the protocluster is the source of the 25\microns emission. The radiation is escaping via the lower density regions created by the large-scale outflow. This shows that even in the densest region of our Galaxy, infrared radiation may find a way to escape instead of being trapped inside and repeatedly scattered 
\citep{thompson_radiation_2005,krumholz_dynamics_2009,menon_infrared_2022,menon_outflows_2023}.
\rrr{Radiation escape in the context of cluster formation and cluster-driven outflows is a more recent concept \citep{peters2012} compared to the much better studied radiation flow around individual protostars \citep[e.g.][]{zhang2013}. 
The radiation escape observed in Sgr B2 N has not previously been seen at such high densities or at such an early stage in cluster formation history \citep[$\sim 0.74$ Myr;][]{kruijssen_dynamical_2015}.
}

This extended emission is present in \rr{SOFIA 37\microns and} \textit{Spitzer} \rr{25\microns} observations, however it is \rr{indistinguishable from} the emission from the surrounding \hii regions \rr{due to lower resolution and lack of multi-wavelength detections, and thus its nature could not be established until now}.
\rr{Future MIRI spectroscopic observations of this emission have the potential to test recent models of UV-radiation-driven, mass-regulating outflows at the extreme densities of early-stage proto-super star clusters \citep[$\Sigma($Sgr B2 N$)\gtrsim10^4 \msun \mathrm{pc}^{-2}$ on parsec scales,][]{menon_outflows_2023}.}



\subsection{\hii regions}
The extended Sgr B2 cloud hosts a large number of \hii regions with scales ranging from 0.005 \pc to 1 \pc \rr{as detected by their cm radio emission} \citep{meng_physical_2022}. Our JWST observations are sensitive to two components of the emission from \hii regions: recombination line emission and thermal emission from hot dust. 
The \rr{central star of the} \hii regions heats the surrounding gas and dust, resulting \rr{in mid-IR emission} seen in MIRI and the long-wavelength NIRCam filters. 
We use the presence of recombination line emission in combination with the extended \rr{mid-IR emission} to identify new \hii region candidates and compare these detections with the existing catalogs of \hii regions \citep[e.g.][]{schmiedeke_physical_2016, meng_physical_2022}. 
\rr{The left panel of Figure \ref{fig:HII_cand_asymmetry} shows the spatial distribution of the known \hii regions as orange circles, and the new JWST-detected \hii region candidates with cyan circles}.



\subsubsection{Deficit at 12\microns relative to common extinction curves}\label{sec:12_deficit}
\rrr{In Section \ref{sec:SED} we constructed a simple model consisting of stellar, dust, and free-free emission to fit the observed SED of X8.33 UC\hii region. A commonly used extinction law \citep{chiar_pixie_2006} was applied with the final fit shown as a red line in Figure \ref{fig:X833}. 
}
The \rrr{observed} SED shows a significant lack of emission at 12.8\microns \rrr{and a deficit at 7.7\microns relative to the fitted SED. 
Since the dust emission is the dominant component of the emission in MIRI filters for this source, the observed deficit is most likely attributed to the extinction model not fully capturing the complexity in the Galactic center.
} 
\rr{The known \water libration mode absorption feature between 10 and 13\microns is likely responsible for the significant dip in the observed flux \citep{Peeters2002,yang_corinos_2022,vanDishoeck2025}, not accounted for in the utilized extinction curve \citep{chiar_pixie_2006}.} This \water libration feature can extend as far as 30\microns.
\rr{In addition,} broad silicate absorption features have been observed in UC\hii regions in the massive star forming region W51A \citep{barbosa_mid-infrared_2016}, extending into the F1280W filter bandwidth.
In combination, these absorption features might be responsible for the deficit seen in MIRI filters\rr{, when compared to common extinction curves towards the Galactic center \rrr{such as the red curve in Figure \ref{fig:X833}}}.

Among the known \rr{radio-detected} \hii regions \rr{that are also detected in the presented JWST observations, a small fraction \rrr{is suitable for the SED analysis described in Section \ref{sec:SED}.} Many of the \hii regions are highly clustered, have complex morphology, are saturated in MIRI filters, or lack 4\microns NIRCam detections. In addition to X8.33, we find the same \rrr{non-detections consistent with significant absorption} at 12.8\microns for all five other \hii regions suitable for the SED analysis: X, Z10.24, L13.30, Y, and A2.}


\subsubsection{JWST is able to detect UC\hii regions missed by radio observations in highly extinguished regions}

\begin{figure*}
    \centering
    \includegraphics[]{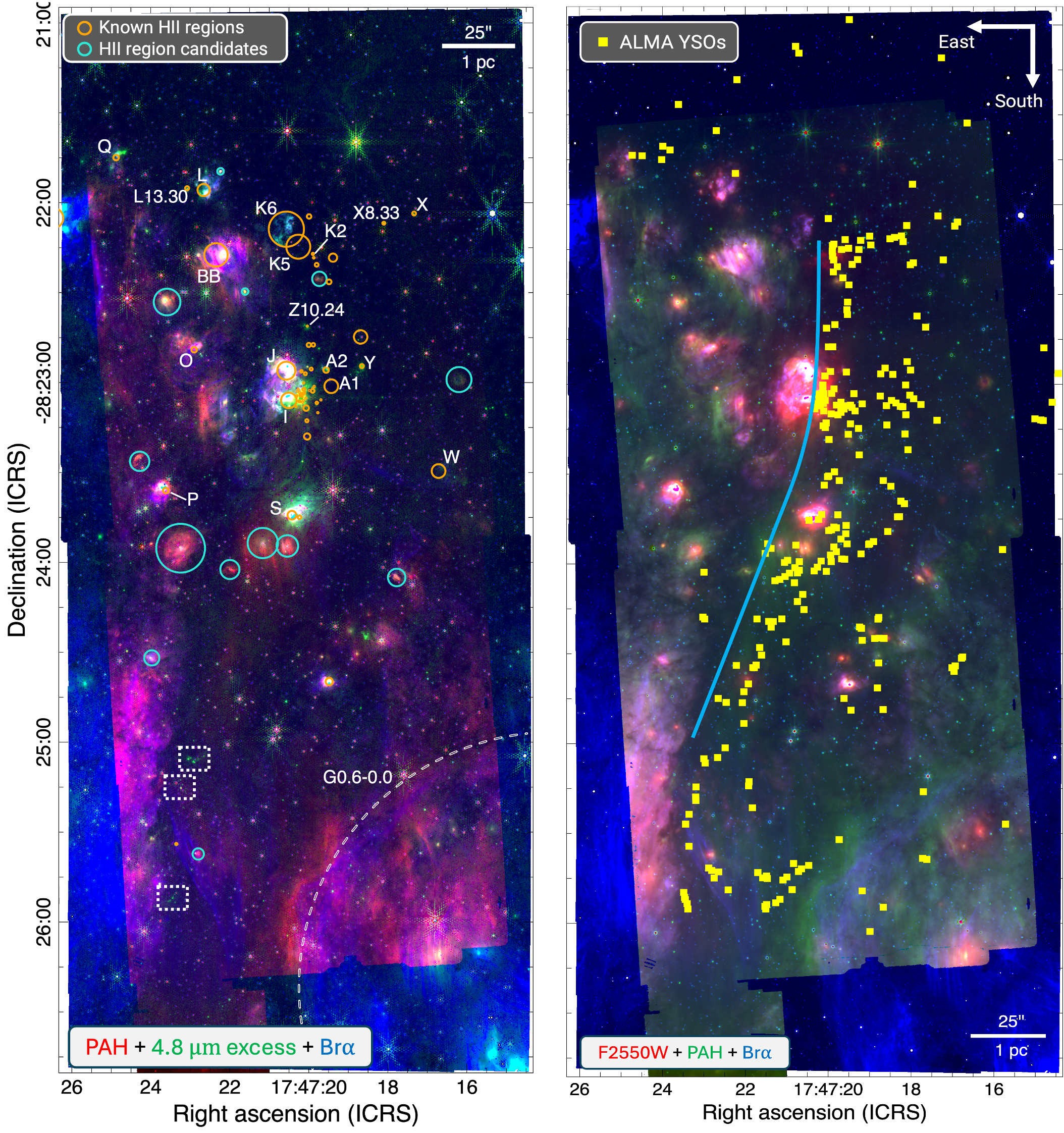}
    \caption{\rr{Left:} A tricolor NIRCam and MIRI image showing the locations of known \hii regions \rr{and new candidate \hii regions detected with JWST} with orange \rr{and cyan} circles \rr{respectively}. PAH emission captured in F770W is shown in red, the contribution from hot dust is shown in green F480M$-[$F410M$-$F405N], and the \bra recombination line is shown in blue. There are \rr{13} \hii region candidates with varying sizes \rr{highlighted with cyan circles, which are drawn to be slightly larger than the apparent size of the \hii region candidates to show the structure of the emission. }
    Several of the extended \hii region candidates are detected with MeerKAT, but \rr{could} not be easily identified due to confusion. The white dotted squares show the locations of zoom-ins in Figure \ref{fig:YSO_cand}.
    \rr{Right:} Star formation asymmetry in Sgr B2: ALMA-detected YSOs are \rr{present} on one side of the cloud \citep{ginsburg_distributed_2018}, while the JWST reveals recently formed stars on the other side of the cloud. The \rr{cyan} line separates the two distinct regions, highlighting the sharp transition. F2550W is shown in red, F770W (PAHs) is shown in green, and \bra recombination line is shown in blue. \rf{We note that while more dusty YSOs have been identified with follow-up observations \citep[][A. Daley, in preparation]{budaiev_protostellar_2024} in this Figure we choose to maintain uniform catalog sensitivity across the whole cloud.}
    }
    \label{fig:HII_cand_asymmetry}
\end{figure*}


We use the presence of \bra emission, polycyclic aromatic hydrocarbon (PAH) emission, and the presence of warm dust to identify \hii region candidates. We identify 13 previously missed sources that exhibit all three features; we designate such sources as \hii region candidates.
\rr{The left panel of} Figure \ref{fig:HII_cand_asymmetry} shows the spatial distribution of VLA-detected \hii regions \citep{schmiedeke_physical_2016, meng_physical_2022} and the \hii region candidates revealed with JWST. 
\rr{An important caveat is that the chosen selection criteria are not unique to \hii regions; planetary nebulae can also exhibit PAH emission, recombination line emission, and warm dust continuum \citep{Anderson2012}. The turbulent environment and complex dynamics of the CMZ result in evolved interlopers distributed across the Galactic center \citep{Dong2015}.  
Thus, some of the sources classified as new \hii region candidates could instead be evolved intermediate mass stars ejected from nearby star clusters.} 

There are two types of \hii region candidates observed in the JWST data: extended and near-unresolved. Comparing with the MeerKAT 1.2\ghz data, we find that some of the extended, VLA-missed \hii regions are detected. However, others are located in regions of extended synchrotron emission, which obscures them from clear identification. We attribute the VLA non-detections to a lack of surface-brightness sensitivity required for the very diffuse \hii regions.

The more compact \hii regions are too small and too close to extended emission to be identified in MeerKAT observations. 
We inspect the \rrr{6}\ghz data from \cite{meng_physical_2022} and identify that these new \hii candidates are either located in the regions significantly impacted by imaging artifacts or their \rrr{6}\ghz flux \rrr{is low and was not identified as a valid source during the cataloging process.}



Both unresolved and slightly extended \hii region candidates in the JWST data exhibit excess in F480M data. We attribute this to the radiation leaking from the UC\hii region and heating up the surrounding dust, thus revealing it in our infrared observations. 

Despite the numerous multiwavelength and multiscale investigations of Sgr B2 cloud, there are still new forming stars being discovered, including massive ones. 
Thus, it is likely that the star formation rate in the region \citep{ginsburg_distributed_2018, budaiev_protostellar_2024} and possibly the whole CMZ \citep[see][]{henshaw_star_2022} is underestimated.


\subsection{ALMA-detected YSOs are not seen with JWST}
We cross-match the protostellar cores seen in ALMA millimeter data 
with the JWST NIRCam catalog. We augment the search with a by-eye inspection of the locations of the ALMA-detected \rrr{sources}.
\rrr{The mm sources in the extended molecular have been cataloged in \cite{ginsburg_high_2018} with a higher resolution follow-ups towards the denser regions of the cloud in \cite[][A. Daley, in preparation]{budaiev_protostellar_2024}. Based on $\sim500$ au resolution observations at 1 and 3 mm, \cite{budaiev_protostellar_2024} determined that the most of the mm detections are likely rotationally supported Stage 0/I YSOs. The free-fall time for the relatively massive sources is too short for all of them to be under gravitational collapse and thus be considered pre-stellar cores. While the central stars for non-\hii region mm-detections have not been directly observed, \cite{xu2025} also suggest that most of these detections are likely YSOs due to their observed sizes and brightnesses (see their section 6.2).
}

\begin{figure*}
    \centering
    \includegraphics[width=1\linewidth]{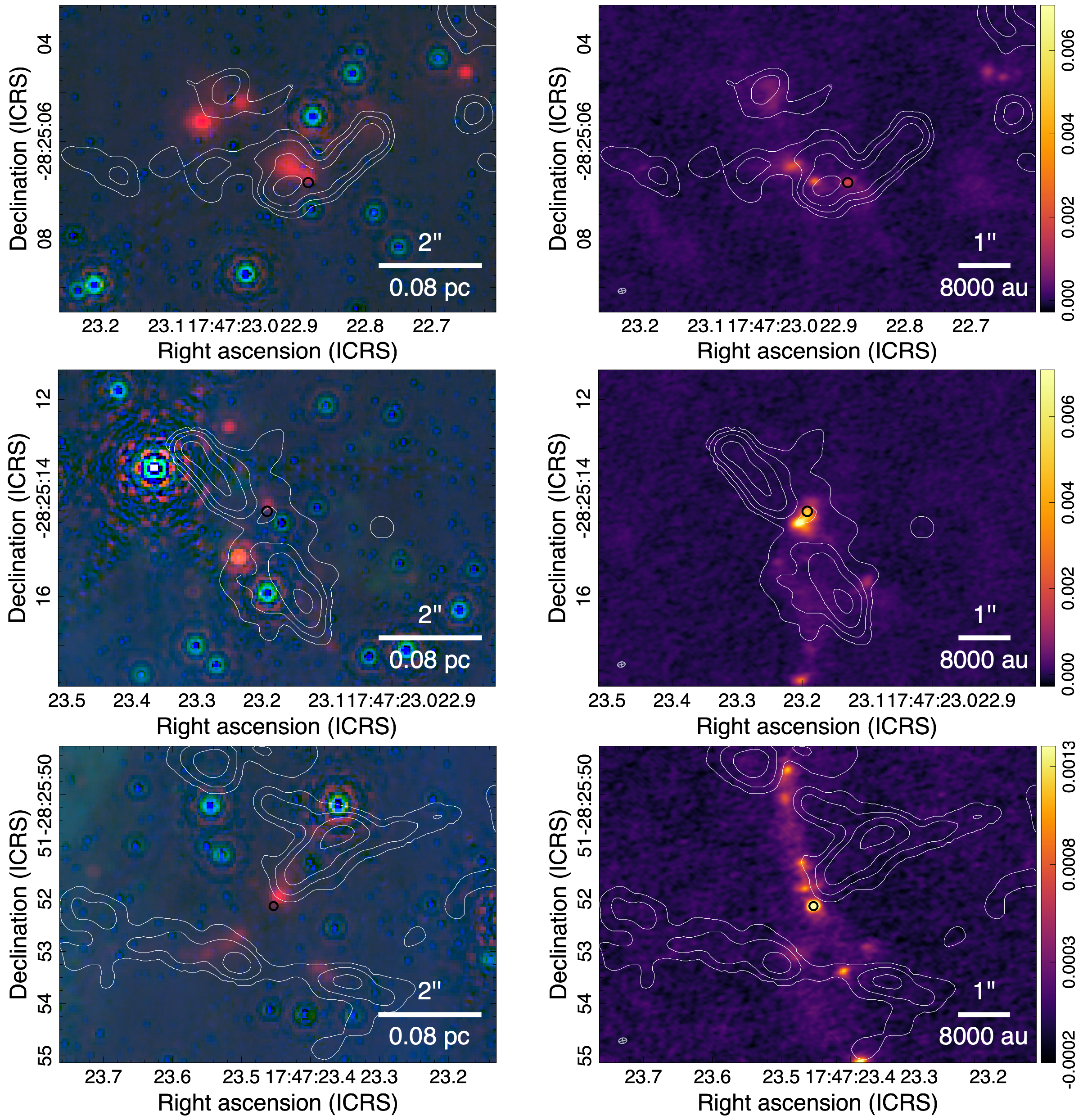}
    \caption{Left column: two selected locations that show 4.8\microns excess that spatially overlaps with sites of star formation activity seen with ALMA. 
    The NIRCam tricolor image has the same colors as Figure \ref{fig:overview_labels}. The relative position of the zoom-ins is shown in \rr{the left panel of} Figure \ref{fig:HII_cand_asymmetry} as white dotted squares. SiO integrated intensity map is shown with white contours indicating the locations of shocked gas produced by outflows. The emission in F480M is slightly extended, indicating that it might be emitted by \rr{either} hot dust continuum \rr{or by the CO gas in the outflow}. \rf{Due to the different PSF shape in medium and narrow bandwidth JWST filters, point sources appear to have rings and dots in a PSF-like patter around them}. 
    Right column: same locations, but as seen at 1 mm continuum. The ALMA continuum does not exactly match the locations of emission in JWST. This suggests that JWST is detecting more diffuse, hot dust that is optically thin at mm wavelengths, while ALMA is detecting the more dense, earlier stages of star formation in the same region that are too \rr{extinguished} to be detected in infrared light. 
    \rr{The \rf{black} circle marks the location of an ALMA-detected YSO that was matched with a JWST detection.}}
    \label{fig:YSO_cand}
\end{figure*}

The high stellar density 
results in line-of-sight, by chance matches with ALMA data. An ALMA detection of a YSO requires the presence of significant amount of dust and, thus, such objects should exhibit strong reddening.
To remove chance line-of-sight matches, we exclude spatial matches without 4.8\microns excess. 
\rr{We define 4.8\microns excess as a significant deviation in the color (F480M $-$ [F410M $-$ F405N]) from the value expected based on extinction, as determined from other filter pairs or, in the case of non-detections, lower limits on the extinction.}
Out of over 700 protostellar cores, only three plausible matches remain, all of which are in Sgr B2 DS. 
The offset between the peak of the 1 mm continuum emission and long-wavelength NIRCam filters is inconsistent between the three sources, ranging between 800 and 2000 au, much larger than the astrometric accuracy $\sim100$ au. 
\rr{Assuming that the mm and mid-IR observations are tracing the same source rather than being a by-chance aligned detection, the different offset vector between mm and IR observations for each of the three object suggests that ALMA and JWST are tracing different parts of the YSOs.
}
\rrr{Two out of the three sources have extended emission nearby in the JWST images, which we discuss in the following subsection.}
\rr{The locations of the \rrr{three} ALMA sources possibly detected with JWST are marked with cyan circles in Figure \ref{fig:YSO_cand}.}

To estimate the required extinction to obscure the vast majority of the ALMA-detected protostellar cores, we use two approaches: extinction of the central stellar object and extinction of the surrounding hot dust.



We select the brightest ALMA-detected source from the Sgr B2 N and M sample that is not affected by saturation or extended emission in the F480M filter: source 168 in the 3 mm catalog from \cite{budaiev_protostellar_2024}. The source is marginally resolved at 1 mm with $\sim500$ au resolution and has a spectral index of 1.87$\pm0.03$ between 1 and 3 mm, indicating optically thick dust. The 22\ghz upper limit measurement is more than an order of magnitude below an extrapolated emission with a slope of 2 \rrr{based on the 3 mm flux}. The 22\ghz non-detection suggest a lack of free-free emission and that the dust becomes optically thin between 3 mm and 1.3 cm.
Assuming a 100\% beam-filling factor of the $r=350$ au beam at 3 mm, the observed radio emission requires $T_d = 170$ K. Following the SED-modeling from the previous subsection and assuming no free-free and stellar contributions, the non-detections at 4.8 and 7.7 \microns require an extinction of $A_V > 200$. \rr{Since t}he majority of the ALMA-detected protostellar cores are located in N(H$_2$)$> 5\times 10^{23}\persc$ \rr($A_V\approx 300$) \citep{ginsburg_distributed_2018}, \rf{the lack of 4.8 and 7.7 \microns detections is not surprising.}

A population of more evolved YSOs would require an even lower extinction to be completely obscured.
\rrr{Assuming a} Stage II/III YSO with a mass of $M = 8\msun$, a higher-end mass of a star that is not producing an \hii region\rrr{: w}ith \rf{a conservative} $M_V\sim-$\rf{2} \citep{Wichittanakom2020}, and $$m_0 = M + 5\log_{10}(d/10\pc),$$ the apparent magnitude is $m_0 = $\rf{12.6}. The F480M observations are able to detect objects down to 21 mag. Converting $A_{4.8}$ = \rf{8.4} to extinction in V band assuming the \cite{chiar_pixie_2006} extinction law results in $A_V \sim $\rf{150}, even lower than the extinction required for the hot dust.
\rr{The distance uncertainty to Sgr B2 of $\sim100$ pc introduces a $<1\%$ error on the measurements described above.}

\subsection{JWST detects \rr{emission adjacent to ALMA YSOs}}
We observe an apparent NIRCam long wavelength excess in some parts of Sgr B2 DS associated with early stages of star formation. Figure \ref{fig:YSO_cand} shows the F480M excess \rrr{in red} next to SiO outflows in \rrr{three} locations in Sgr B2 DS \rr{with the most apparent association between ALMA and JWST observations}. The locations of the \rrr{three} cutouts within the cloud are shown in Figure \ref{fig:HII_cand_asymmetry} as white dotted \rrr{rectangles}. Based on the close spatial association with regions with known outflows and YSOs we assume that this emission is the product of the active star formation. Considering the linearity of the \rrr{extended} features, we hypothesize that NIRCam emission is coming from the hot dust around the outflow cavities. \rr{Alternatively, the F480M excess could be attributed to CO bandhead emission from the outflow \citep{Ray2023, Hodapp2026}.}

\rr{Just over a dozen} similar objects with F480M excess, usually unresolved, are found throughout the cloud, but they are not distributed uniformly. Some of these sources are present in locations with no known star-forming activities \rr{based on radio observations}. One explanation is that as the YSOs evolve \rr{and become undetectable at mm wavelengths}, they move outwards within the cloud in a manner reminiscent of the observed distribution of YSOs in Orion \citep{Grossschedl2019}. Alternatively, it is possible that these (some or all) are very reddened stars. 


\subsection{Cloud morphology}


\subsubsection{Star formation asymmetry seen in ALMA and JWST}
Over the years of radio observations of Sgr B2, it has become apparent that the majority of the star formation in the cloud is happening towards its western side (\rr{leftward} in the figures)  relative to 
\rr{Sgr B2 N and M} protoclusters \citep[e.g.][]{schmiedeke_physical_2016, ginsburg_distributed_2018, meng_physical_2022}. Studies of protostellar cores and \hii regions, as well as high-resolution and high-sensitivity follow-ups show that there is ongoing star formation in the Sgr B2 N, M, and S protoclusters with ``chains" of YSOs to one side and an absence of dusty cores on the other side. Figure \ref{fig:HII_cand_asymmetry} shows that the number of radio-detected YSOs and \hii regions declines slowly in one direction, while forming a sharp edge in the other direction. 

JWST observations paint a different picture: most of the recently formed stars are located opposite to those seen in ALMA. The protoclusters are only barely peeking through or are detected indirectly in case of Sgr B2 N, as discussed in Section \ref{sec:MIRI_SgrB2_N}. Figure \ref{fig:HII_cand_asymmetry} further highlights that the ionized and heated gas from the recently formed stars is preferentially detected on the eastern side of the cloud.
The apparent anti-correlation between mm and infrared emission can be explained with a density asymmetry in the cloud: the low-density, eastern side of the cloud does not have sufficient mass to form new dusty YSOs, and the recently formed stars illuminate the diffuse medium that is not dense enough to be detected in radio observations. On the other hand, the western, dense side of the cloud has many accreting, dusty YSOs and is completely \rr{extinguished} in infrared wavelengths. 
\rrr{It is uncertain how this sharp-edged asymmetry formed.}
It is also unclear whether the young stars in the eastern side of the cloud formed there and eventually cleared out the surrounding material or if they formed in the western side of the cloud and moved eastward over time. 

We also note that the YSO ``chains" and mm continuum filamentary structures seen outside of Sgr B2 N, M, and S in the western side of the cloud \citep{budaiev_protostellar_2024} are preferentially oriented in the east-west direction. Such an orientation may be indicative of either the star-forming objects moving through the dense cloud or the dense surrounding material passing through the region of ongoing star formation.


\subsubsection{Sgr B2 has a sharp edge}\label{sec:sharp_edge}
\begin{figure*}
    \centering
    \includegraphics[width=1\linewidth]{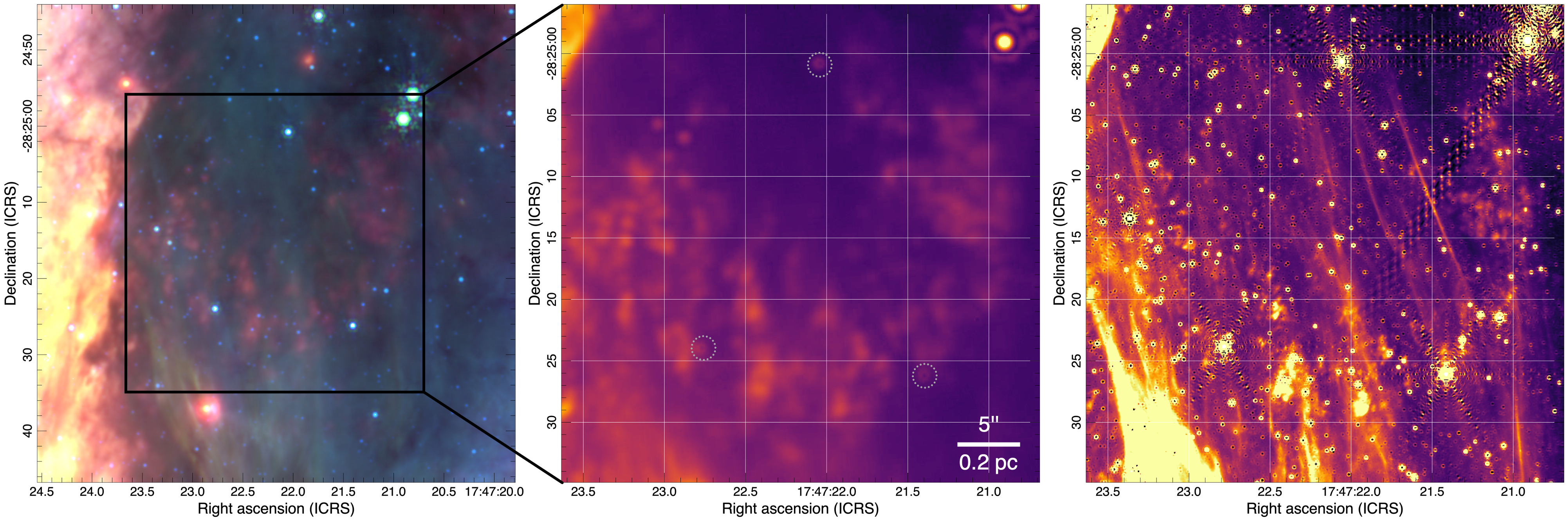}
    \caption{Left: MIRI tricolor image covering part of Sgr B2 DS that shows extended flocculent emission. The structure is only apparent at 25\microns due to other large-scale emission at other wavelengths. Middle: F2550W-only view of the emission. The feature appears to have a spherical shape with dense material extincting some of the emission \rr{towards the top of the image}. The cause for this flocculent structure as well as the source of the emission remain unclear. \rrr{Dashed \rf{gray} circles mark the location of sources affected by persistence in F2550W (see Appendix \ref{sec:persistence}.)} Right: same region as the middle panel, but as seen in \bra.}
    \label{fig:fluff}
\end{figure*}

We observe evidence of the sharp eastern edge of the cloud in Sgr B2 DS (see the \rr{lower} left side of Figures \ref{fig:overview_labels}, \ref{fig:HII_cand_asymmetry}, and \rr{left side of Figure} \ref{fig:fluff}). The extended background becomes rapidly \rr{extinguished} by the dense cloud as we move into Sgr B2 DS. To our knowledge, this is the \rr{most visually defined} cloud edge
in a molecular cloud.
\rr{The \rrr{formation} origin of this sharp edge, reminiscent of ionization fronts in nebulae, is unclear. }
The observed cloud edge in Sgr B2 DS is consistent with the edge of SiO J=2$-$1 cavity D reported by \cite{armijos-abendano_structure_2020}.

\subsubsection{G0.6-0.0 is in front of Sgr B2}
The NIRCam footprint includes an extended \hii region G0.6-0.0, which is part of the ridge of features connecting Sgr B2 and Sgr B1. 
\rr{The relative line-of-sight location of Sgr B2 and Sgr B1 has not been firmly established, with some placing Sgr B2 behind Sgr B1 \citep{Bieging1980, Ryu2009, sofue_bow-shock_2024}, while others arguing that Sgr B2 is located in the foreground to the rest of Sgr B complex \citep{mehringer_radio_1992, mehringer_radio_1993, harris_sofia-upgreat_2021}.}
Furthermore, the widely adopted orbital configuration over the past decade is such that the younger cloud, Sgr B2, is located in the front, with G0.6-0.0 and the older Sgr B1 cloud further along the back-side of the orbit \citep{kruijssen_dynamical_2015, barnes_star_2017}.
However, \rr{our} NIRCam observations suggest that G0.6-0.0 could be located in front of Sgr B2. 
As discussed in Section \ref{sec:sharp_edge}, the dense star forming ridge Sgr B2 DS is seen as a dark patch \rr{completely extinguishing} the extended background emission: 
a clear and sharp edge of the dusty cloud can be seen absorbing the background recombination line emission and glowing dust \rr{east of Sgr B2 DS}. 
However, where Sgr B2 DS meets with the footprint of G0.6-0.0 \hii region in the bottom \rr{right} of the images\rr{, west of Sgr B2 DS,} no such sharp features are seen. 
\rr{Instead, the recombination line emission extends until the boundary of the photodissociation region and the PAH emission extends even further, overlapping with the projected location of Sgr B2 DS \citep{meng_physical_2019}, as seen in F770W filter in Figure \ref{fig:HII_cand_asymmetry} \rrr{when compared to white contours in Figure \ref{fig:overview_labels}.}}
\rrr{Furthermore, dusty YSOs associated with Sgr B2 DS are found in the plane-of-sky location of G0.6-0.0, shown as yellow squares in the right panel of Figure \ref{fig:HII_cand_asymmetry}. Follow-up high sensitivity observations of Sgr B2 DS show even more dusty YSOs at the location of G0.6-0.0 (A. Daley in prep.).}
The lack of \rr{spatial} absorption features strongly suggests that G0.6-0.0 is not \rr{being extincted} by Sgr B2 DS.

\rf{Therefore, based on the data presented in this paper, we argue that G0.6-0.0 is likely located in front of Sgr B2 DS.}
The recombination line-based extinction to parts of G0.6-0.0 is $A_V \sim 50$, indicating that the \hii region is not located at the ``surface" of the CMZ either.
Such spatial configuration is consistent with the bow-shock model of Sgr B cloud complex proposed by \cite{sofue_bow-shock_2024}. However, no large-scale structures validating the presence of a bow shock are present, likely due to the compact spatial coverage. 
\rr{Since G0.6-0.0 is physically related to Sgr B1 region \citep{Mehringer1992}, it's foreground location relative to Sgr B2 DS places the line-of-sight location of Sgr B1 complex in question.}

\subsubsection{Extended flocculence in Sgr B2 DS}
We observe extended 25\microns emission in part of Sgr B2 DS. The emission is flocculent, with identifiable local peaks and missing patches as seen in Figure \ref{fig:fluff}. 
Parts of the emitting region are visible in F1280W, F770\rr{, and \bra recombination line}, however the emission is extremely faint and undistinguishable among other large-scale features. 
\rr{This emission is not detected with neither SOFIA at 25 and 37\microns nor \textit{Spitzer} at 24\microns likely due to the combination sensitivity and resolution.}

The patchy emission begins near the edge of the Sgr B2 DS region and extends for 30\arcsec, forming a circular shape. The top side of the region is dimmer, either due to lack of the underlying emission, or because of foreground material obscuring the emission. We do not know what drives this unique morphology. While the rough symmetry seen in Figure \ref{fig:fluff} (right) hints at a central feedback-driven mechanism, there is no obvious configuration that would produce such a flocculent morphology. 


\begin{figure*}[htbp]
    \centering

    \begin{subfigure}[t]{0.49\textwidth}
        \centering
        \includegraphics[width=\linewidth]{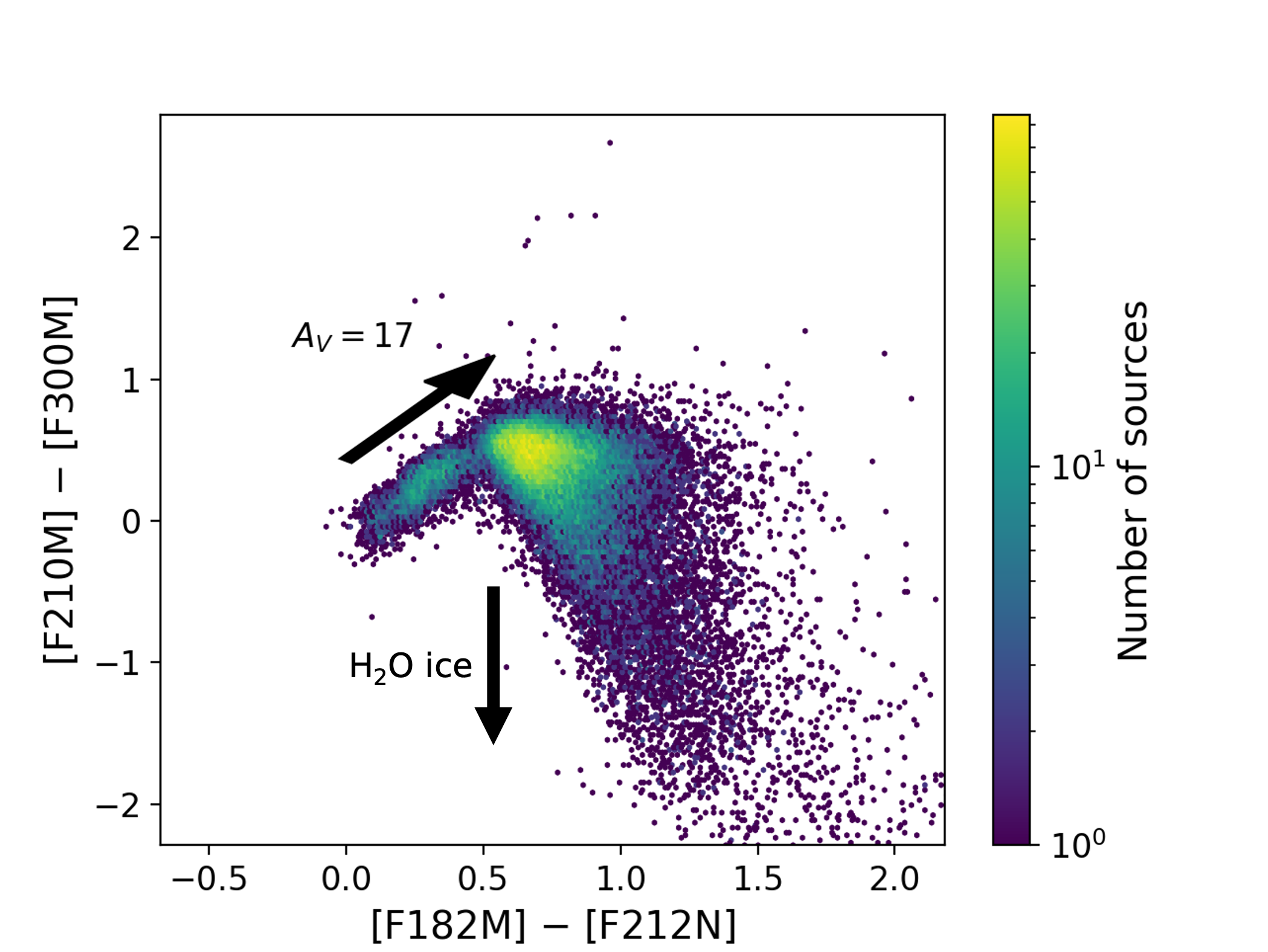}
        \caption{A CCD showing a significant absorption in F300M filter from \water ice.}
    \end{subfigure}
    \hfill
    \begin{subfigure}[t]{0.49\textwidth}
        \centering
        \includegraphics[width=\linewidth]{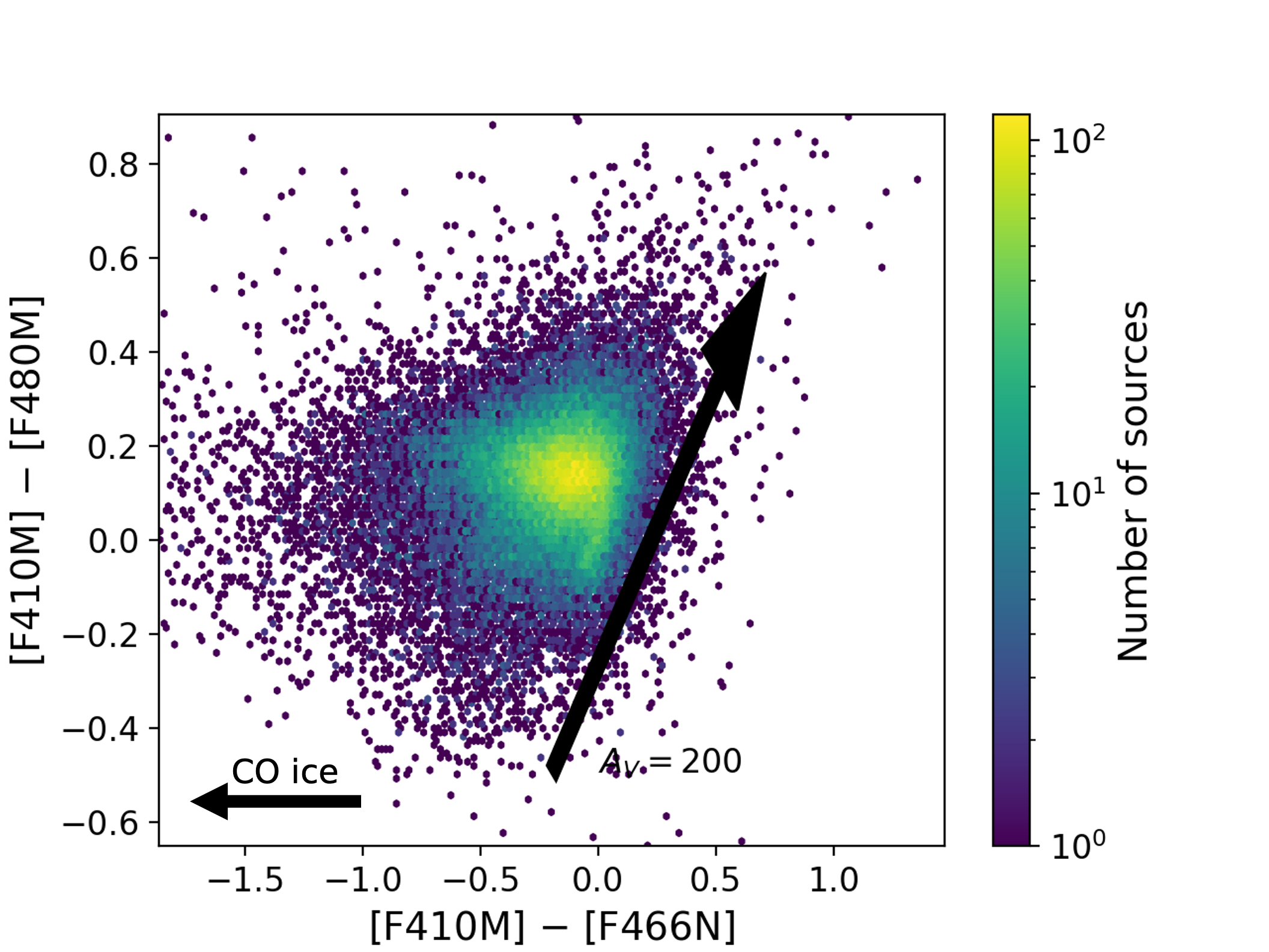}
        \caption{A CCD showing a significant absorption in F466N filter from CO ice.}
    \end{subfigure}

    \vspace{0.5em}

    \begin{subfigure}[t]{0.49\textwidth}
        \centering
        \includegraphics[width=\linewidth]{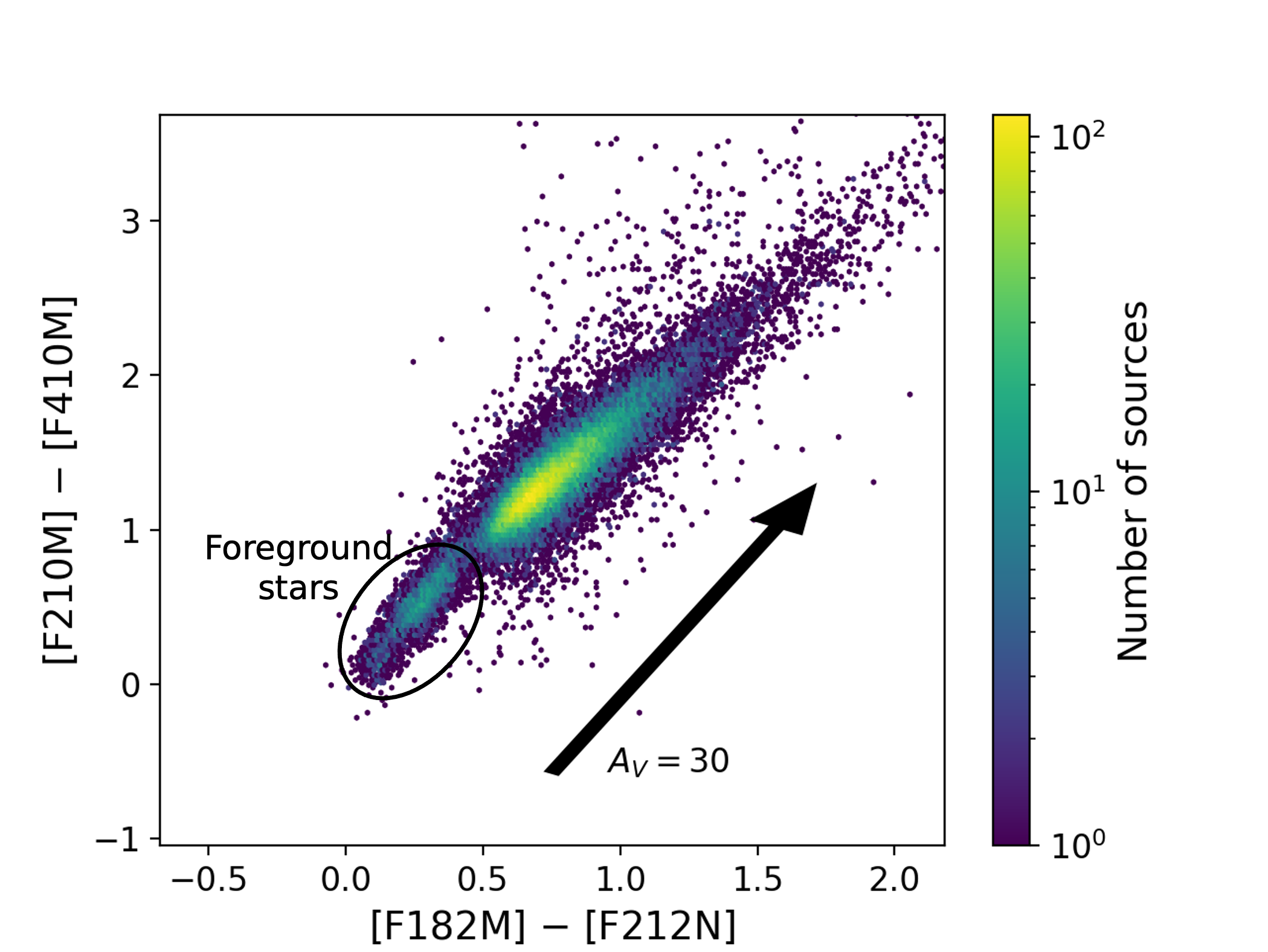}
        \caption{A CCD showing that the observed stellar extinction generally aligns with the extinction model from \cite{chiar_pixie_2006}.}
    \end{subfigure}
    \hfill

    \caption{Color-color and color-magnitude diagrams showing the general features seen in the cloud. Significant amount of \water and CO ice is present in the cloud. }
    \label{fig:diagrams}
\end{figure*}

\subsection{Preliminary CCDs}\label{sec:ccds}
We use \rr{our} preliminary NIRCam and MIRI \rrr{source} catalogs to produce color-color diagrams (CCDs) and color-magnitude diagrams. 
We use the version of the catalog where any detection satisfies the filtering criteria in all filters where it is detected \rr{described in Section \ref{sec:source_extraction}, totaling 84,490 sources}.

As expected for a dense molecular cloud, a significant amount of ice is present. The top row of Figure \ref{fig:diagrams} highlights \water ice absorption in F300M and CO ice absorption in F466N. 
\rr{Stars foreground to the CMZ are seen in panels (a) and (c) as a distinct group of sources that follow the extinction vector up to $A_V \approx 17$.}
\rr{The apparent foreground extinction of 17 magnitudes is consistent with previous measurements of the CMZ and nearby clouds \citep{nogueras-lara_galacticnucleus_2021, Ginsburg2025}. 
In panel (a), the sources in the CMZ show significant absorption from \water ice: without the presence of ice, the stars follow the extinction vector until they hit the CMZ where in the presence of ice absorption in F300M the stars shift downwards on the CCD.}
Panel (b) shows a similar effect from the CO ice\rr{: the absorption in F466N shifts the sources leftward on the CCD}. 
CCDs with the F360M filter show slight lack of 3.6\microns emission at high extinction. We attribute this to the broad 3\microns \water or methanol absorption affecting the F360M filter.
Panel (c) shows that [F182M]$-$[F212N] and [F212N]$-$[F410M] generally line up with the extinction vector with a slight contribution from CO ice at the edges of the F410M filter. 



\section{Conclusions}\label{sec:conclusions}
We presented JWST's NIRCam and MIRI first view of the most vigorously star forming cloud in the CMZ, Sagittarius B2. 
\begin{itemize}
    \item The constructed SEDs of UC\hii regions show significant deficit at 12.8\microns. The \rr{strong \water ice libration and} broadening of the 10\microns silicate absorption feature in a dense environment as well as a strong water libration feature are the likely causes.
    \item MIRI revealed infrared radiation escaping the dense protocluster Sgr B2 N following the path of a large-scale outflow. The presence of extended 25\microns emission suggests that infrared photons do not get trapped within dense clusters. 
    \item Despite extensive radio regime studies of the region, we detect over a dozen new \hii region candidates in the cloud. The high sensitivity of \rr{\bra recombination line} observations allowed for detection of diffuse \hii regions.
    \item ALMA-detected YSOs are too deeply embedded to be detected with JWST. However, JWST appears to detect hot dust around the outflow cavities. 
    
    \item The cloud shows very structured morphology. The earliest stages of star formation are present in the western side of the cloud as seen with ALMA, while JWST detects recent star formation in the eastern, less dense part of the cloud. 
    \rr{The extended recombination line emission highlights the sharp eastern edge of the cloud.} 
    \rr{G0.6-0.0, which is an \hii region thought to be associated with the Sgr B complex, appears to be in front of Sgr B2 DS. This is contrary to some models that place G0.6-0.0 close to Sgr B1 in the back side of the CMZ's orbit.}
\end{itemize}

\begin{acknowledgments}
We thank Dani Lipman and Sheila Sagear for helpful discussions. 
This work is based on observations made with the NASA/ESA/CSA James Webb Space Telescope. The data were obtained from the Mikulski Archive for Space Telescopes at the Space Telescope Science Institute, which is operated by the Association of Universities for Research in Astronomy, Inc., under NASA contract NAS 5-03127 for JWST. These observations are associated with program \#5365.
NB and AG acknowledge support from the Space Telescope Science Institute via grant No. JWST-GO-05365.001-A. 
AG acknowledges support from the NSF under grants AAG 2206511 and CAREER 2142300.
TY acknowledges support from the Space Telescope Science Institute via grant No. JWST-GO-06151.001-A.
SG and AG acknowledge support from the Space Telescope Science Institute via grant No. JWST-GO-02221.001-A. 
E.A.C.\ Mills  gratefully  acknowledges  funding  from the National  Science  Foundation  under  Award  Nos. 1813765, 2115428, 2206509, and CAREER 2339670. 
X.L.\ acknowledges support from the Strategic Priority Research Program of the Chinese Academy of Sciences (CAS) Grant No.\ XDB0800300, the National Key R\&D Program of China (No.\ 2022YFA1603101), State Key Laboratory of Radio Astronomy and Technology, the National Natural Science Foundation of China (NSFC) through grant Nos.\ 12273090 and 12322305, the Natural Science Foundation of Shanghai (No.\ 23ZR1482100), and the CAS ``Light of West China'' Program No.\ xbzg-zdsys-202212.

The authors acknowledge University of Florida Research Computing for providing computational resources and support that have contributed to the research results reported in this publication. URL: http://www.rc.ufl.edu.
\end{acknowledgments}

\begin{contribution}

N.B. acquired the observations, led the analysis, writing, and interpretation. A.G. acquired the observations, oversaw the project progress, 
and contributed to the interpretation and writing. A.T.B. and D.J. were instrumental in acquiring the observations and contributed to the interpretation. T.Y. contributed to the interpretation. C.B., A.B., X.L., E.A.C.M., and D.L.W. were instrumental in acquiring the observations. S.G. contributed to data reduction.

\end{contribution}

\facilities{JWST, ALMA}

\software{astropy \citep{astropy_collaboration_astropy_2013, astropy_collaboration_astropy_2018, astropy_collaboration_astropy_2022},  
          }

\appendix
\section{Persistence in MIRI}\label{sec:persistence}
The F2550W data show depressed central pixels for some of the sources \rrr{as highlighted by dashed circles in \rf{the middle panel of} Figure \ref{fig:fluff}.} This is especially apparent in the dimmer point sources. The depressed pixels are present in the raw, uncalibrated files across all five dithers. This is indicative of persistence caused by the prior observations with the F1280W filter. This is further evidenced by the presence of depressed pixels in the combined image at the locations where the bright star was on the detector at each of the dithers. Points sources above $\sim$0.001\jy in the F1280W filter are affected by persistence. The effect is especially amplified in the sources that are bright in F1280W while being dimmer in F2550W. 
Both F2550W and F1280W filters have several regions with slew persistence, another typical artifact for fields with bright stars. 

\cite{dicken_jwst_2024} reports $<0.01\%$ flux contribution from persistence after 15 minutes of the saturating observations. At the same time, they report an instance of slew persistence that lasted much longer than other examples and modeling, and no cause was identified (PID 2736). While the commissioning investigations of persistence did not include the evaluation of the impact on the ramp fitting process, the persistence-induced error is usually estimated to be below 1\%. However, we observe that the central few pixels are dimmed by up to 5\% relative to the surrounding pixels. 

Currently, no persistence mitigation tools are implemented in the STScI's pipeline for MIRI observations. Nevertheless, it is likely possible to use the depressed pixels at the locations without stellar contribution to model the persistence and apply the correction to the affected pixels.



\bibliography{references-2}{}
\bibliographystyle{aasjournalv7}

\end{document}

%% file: LaTeX-macros.tex
\newcommand{\y}[1]{\ensuremath{y_{#1}}\xspace}

\newcommand{\kms}{\ensuremath{\,{\rm km\,s^{-1}}}\xspace}

\newcommand{\microns}{\ensuremath{\,\mu{\rm m}}\xspace}
\newcommand{\m}{\ensuremath{\,{\rm m}}\xspace}

\newcommand{\pc}{\ensuremath{\,{\rm pc}}\xspace}
\newcommand{\kpc}{\ensuremath{\,{\rm kpc}}\xspace}
\newcommand{\K}{\ensuremath{\,{\rm K}}\xspace}

\newcommand{\yr}{\ensuremath{\,{\rm yr}}\xspace}

\newcommand{\ghz}{\ensuremath{\,{\rm GHz}}\xspace}

\newcommand{\persc}{\ensuremath{\,{\rm cm^{-2}}}\xspace}

\newcommand{\jy}{\,Jy\xspace}

\newcommand{\Mjysr}{\ensuremath{{\rm \,MJy\,sr^{-1}}}\xspace}
\newcommand{\msun}{\ensuremath{{\rm \,M_\odot}}\xspace}     
\newcommand{\rsun}{\ensuremath{{\rm \,R_\odot}}\xspace}     
 
\newcommand{\h}[1]{\ensuremath{^{#1}{\rm H}}\xspace}

\newcommand{\hii}{{\rm H\,{\small II}}\xspace}

\newcommand{\htwo}{\ensuremath{{\rm H_2}}\xspace}

\newcommand{\paa}{\ensuremath{\mathrm{Pa}\alpha}\xspace}
\newcommand{\bra}{\ensuremath{\mathrm{Br}\alpha}\xspace}

\newcommand{\water}{\ensuremath{\rm H_2O}\xspace}